\def\year{2021}\relax
\documentclass[letterpaper]{article} 
\usepackage{aaai21}  
\usepackage{times}  
\usepackage{helvet} 
\usepackage{courier}  
\usepackage[hyphens]{url}  
\usepackage{graphicx} 
\usepackage{natbib}  
\usepackage{caption} 
\usepackage[inline]{enumitem} \setlist{nosep}
\RequirePackage{xspace}
\usepackage{mathtools}
\usepackage{xcolor, graphicx}
\usepackage{subfig}
\usepackage[english]{babel}
\usepackage{latexsym,bm,bbm,amsmath,amssymb}
\usepackage{mathrsfs}
\usepackage{etoolbox}
\RequirePackage{tcolorbox}
\usepackage{cleveref}
\RequirePackage[skip=.5ex]{caption}
\RequirePackage{newfloat}
\DeclareFloatingEnvironment[fileext=algo,name=Algorithm]{algofig}
\RequirePackage{tikz}
\usetikzlibrary{decorations.pathreplacing}
\usetikzlibrary{arrows}
\RequirePackage{adjustbox}
\usepackage{multirow} 
\usepackage{dsfont,subfig}
\RequirePackage[noend]{algpseudocode}

\RequirePackage[textsize=scriptsize,color=yellow!25,linecolor=brown]{todonotes}

\makeatletter
\g@addto@macro{\normalsize}{%
\setlength{\abovedisplayskip}{3pt plus1pt}%
\setlength{\abovedisplayshortskip}{3pt plus1pt}%
\setlength{\belowdisplayskip}{3pt plus1pt}%
\setlength{\belowdisplayshortskip}{3pt plus1pt}}
\makeatother

\usepackage{Definitions}

\graphicspath{{./FIG/}}

\def\our{\textsc{Perm\-Gnn}\xspace}
\def\google{\text{Google+}\xspace}
\DeclareMathOperator{\sign}{sign}
\DeclareMathOperator{\loss}{loss}
\DeclareMathOperator{\similarity}{sim}
\DeclareMathOperator{\nbr}{nbr}
\DeclareMathOperator{\nnbr}{\overline{\text{nbr}}}
\DeclareMathOperator{\aggregate}{\textsc{Aggr}}
\DeclareMathOperator{\combine}{\textsc{Comb}}
\DeclareMathOperator{\RowScale}{\textsc{RowScale}}
\DeclareMathOperator{\ColScale}{\textsc{ColScale}}
\DeclareMathOperator{\KTau}{\textsc{KTau}}

\newcommand{\set}[1]{\{ #1 \}}
\def\ne{\overline{\Ecal}}

\newcommand{\gap}{\vspace{.5mm plus.1mm}}
\renewcommand{\Gcal}{G}
\renewcommand{\Vcal}{V}
\renewcommand{\Ecal}{E}
\renewcommand{\Qcal}{Q}
\newcommand{\xhdr}[1]{\vspace{0.1mm}\noindent{{\bfseries #1.}}}
\renewcommand{\paragraph}{\xhdr}

\newcommand{\best}[1]{\textbf{#1}}

\def\ztitle{Adversarial Permutation Guided Node Representations for  Link Prediction}
\title{\ztitle}
 
\author { Indradyumna Roy, Abir De,
        Soumen Chakrabarti\\ }
\affiliations {Indian Institute of Technology Bombay\\ \{indraroy15, abir, soumen\}@cse.iitb.ac.in}

\begin{document}

\maketitle

\begin{abstract}
After observing a snapshot of a social network, a link prediction (LP) algorithm identifies node pairs between which new edges will likely materialize in future. Most LP algorithms estimate a score for currently non-neighboring node pairs, and rank them by this score.   Recent LP systems compute this score by comparing dense, low dimensional vector representations of nodes.  Graph neural networks (GNNs), in particular graph convolutional networks (GCNs), are popular examples.  For two nodes to be meaningfully compared, their embeddings should be indifferent to reordering of their neighbors.  GNNs typically use simple, symmetric set aggregators to ensure this property, but this design decision has been shown to produce representations with limited expressive power.  Sequence encoders are more expressive, but are permutation sensitive by design.  Recent efforts to overcome this dilemma turn out to be unsatisfactory for LP tasks. In response, we propose \our, which aggregates neighbor features using a recurrent, order-sensitive aggregator and directly minimizes an LP loss while it is `attacked' by adversarial generator of neighbor permutations.  \our{} has superior expressive power compared to earlier GNNs.  Next, we devise an optimization framework to map \our's node embeddings to a suitable locality-sensitive hash, which speeds up reporting the top-K most likely edges for the LP task. Our experiments on diverse datasets show that \our outperforms several state-of-the-art link predictors, and can predict the most likely edges fast.
\end{abstract}

\section{Introduction}
\label{sec:Intro}

In the link prediction (LP) task, we are given a snapshot of a social network, and asked to predict future links that are most likely to emerge between nodes. LP has a wide variety of applications, \eg, recommending friends in Facebook, followers in Twitter, products in Amazon, or connections on LinkedIn.  An LP algorithm typically considers current non-edges as potential edges, and ranks them by decreasing likelihoods of becoming edges in future.

\subsection{Prior Work and Their Limitations}
\label{sec:PriorWork}
LP methods abound in the literature, and predominantly follow two approaches. The first approach relies strongly on hand-engineering node features and edge likelihoods based on the network structure and domain knowledge~\cite{Katz1997start,LibenNowellK2007LinkPred,BackstromL2011SRW}. However, such feature engineering often demands significant domain expertise. The second approach learns low dimensional node embeddings which serve as node features in LP tasks.  Such embedding models include Node2Vec~\cite{grover2016node2vec}, DeepWalk~\cite{perozzi2014deepwalk}, etc., and various graph neural networks (GNN), \eg, GCN~\cite{kipf2016semi}, GraphSAGE~\cite{hamilton2017inductive}, GAT~\cite{velivckovic2017graph}, etc.


%

\paragraph{Limited expressive power of GNNs}
While deep graph representations have shown significant potential in capturing complex relationships between nodes and their neighborhoods, they lack representational power useful for LP.  A key reason for this weakness is the use of symmetric aggregates over a node $u$'s neighbors, driven by the desideratum that the representation of $u$ should be invariant to a permutation of its neighbor nodes~\citep{zaheer2017deep,ravanbakhsh2016deep,qi2017pointnet}.  Such networks have recently been established as low-pass filters \citep{wu2019simplifying, nt2019revisiting}, which attenuate high frequency signals.  This prevents LP methods based on such node representations from reaching their full potential. 
Although recent efforts \citep{lee2019set, bloem2019probabilistic, shi2020deep, stelznergenerative, skianis2020rep,ZhangC2018LinkPredGNN} on modeling inter-item dependencies have substantially improved the expressiveness of set representations in applications like image and text processing,  they offer only modest improvement for LP, as we shall see in our experiments.
Among these approaches, SEAL \citep{ZhangC2018LinkPredGNN} improves upon GNN performance but does not readily lend itself to efficient top-$K$ predictions via LSH.

\paragraph{Limitations of sequence driven embeddings}
We could arrange the neighbors of $u$ in some arbitrary canonical order, and combine their features sequentially using, say, a recurrent neural network (RNN).  This would capture feature correlations between neighbors.  But now, the representation of $u$ will become sensitive to the order in which neighbors are presented to the RNN.  In our experiments, we see loss degradation when neighbors are shuffled.  We seek to resolve this central dilemma.  An obvious attempted fix would be to present many permutations (as Monte Carlo samples) of neighbor nodes but, as we shall see, doing so in a data-oblivious manner is very inefficient in terms of space and time.

\subsection{Our Proposal: \our}

In response to the above limitations in prior work, we develop \our: a novel node embedding method specifically designed for LP.  To avoid the low-pass nature of GNNs, we eschew symmetric additive aggregation over neighbors of a node $u$, instead using a recurrent network to which neighbor node representations are provided sequentially, in some order.  The representation of $u$ is computed by an output layer applied on the RNN states.

To neutralize the order-sensitivity of the RNN, we cast LP as a novel min-max optimization, equivalent to a game between an adversary that generates worst-case neighbor permutations (to maximize LP loss) and a node representation learner that refines node representations (to minimize LP loss) until they become insensitive to neighborhood permutations.  To facilitate end-to-end training and thus avoiding exploration of huge permutation spaces, the adversarial permutation generator is implemented as a Gumbel-Sinkhorn neural network \citep{Mena+2018GumbelSinkhorn}.


Next, we design a hashing method for efficient LP, using the node representation learnt thus far. We propose a smooth optimization to compress the learned embeddings
into binary representations, subject to certain hash performance
constraints. Then we leverage locality sensitive hashing \citep{GionisIM1999hash} to assign the bit vectors to buckets, such that nodes likely to become neighbors share buckets. Thus, we can limit the computation of pairwise scores to within buckets. In spite of this additional compression, our hashing mechanism is accurate and fast.

We evaluate \our on several real-world datasets, which shows that our embeddings can suitably distill information from node neighborhoods into compact vectors, and offers accuracy boosts beyond several state-of-the-art LP methods\footnote{Code: \url{https://www.cse.iitb.ac.in/~abir/codes/permgnn.zip}.}, while achieving large speed gains via LSH.

\subsection{Summary of Contributions} 
(1)~\textbf{Adversarial permutation guided embeddings:}
We propose \our, a novel node embedding method, which provides high quality node representations for LP.
In a sharp contrast to additive information aggregation in GNNs,
we start with a permutation-sensitive but highly expressive aggregator
of the graph neighbors and then desensitize the permutation-sensitivity
by optimizing a min-max ranking loss function 
with respect to the smooth surrogates of adversarial permutations. 

\noindent (2)~\textbf{Hashing method for scalable predictions:}
We propose an optimized binary transformation to the learnt node representations, that readily admits the use of a locality-sensitive hashing method and shows fast and accurate predictions.

\noindent (3)~\textbf{Comprehensive evaluation:} We provide a rigorous evaluation to test both the representational
power of \our and the proposed hashing method, which show that our proposal usually outperforms classical and recent methods.
Further probing the experimental results reveal insightful explanations behind the success of \our.

\section{Preliminaries}
\label{sec:Prelim}

In this section, we describe necessary notation and the components of a typical LP system.
\subsection{Notation}
We consider a snapshot of an undirected social network $\Gcal=(\Vcal,\Ecal)$. 
Each node $u$ has a feature vector $\fb_u$. 
We use $\nbr(u)$ and $\nnbr(u) $ to indicate the set of neighbors and non-neighbors of $u$.  Our graphs do not have self edges, but we include $u$ in $\nbr(u)$ by convention.  We define $\nbr(u)=\set{u}\cup\set{v\,|\, (u,v)\in \Ecal}$, $\nnbr(u)=\set{v| v \neq u, (u,v)\not \in \Ecal}$ and also $\ne$ to be the set of non-edges, \ie, $\ne=\cup_{u\in\Vcal}\nnbr(u)$. Finally, we define $\Pi_\delta$ to be the set of permutations of the set $[\delta]=\set{1,2,..., \delta}$ and $\Pcal_\delta$ to be the set of all possible 0/1 permutation matrices of size $\delta\times\delta$.
\subsection{Scoring and Ranking}
Given a graph snapshot $\Gcal=(\Vcal,\Ecal)$, the goal of a LP algorithm is to identify node-pairs from the current set of non-edges $\ne$ (often called potential edges) that are likely to become edges in future.
In practice, most LP algorithms compute a \textbf{score} $s(u,v)$ for each potential edge $(u,v)\in\ne$, which measures their likelihood of becoming connected in future.
%
Recently invented network 
embedding methods~\cite{kipf2016semi,grover2016node2vec,Salha+2019gravity}
first learn a latent representation $\xb_u$ of each node $u\in\Vcal$ and then compute scores $s(u,v)$ using some similarity or distance measure between the corresponding representations $\xb_u$ and $\xb_v$.
In the test fold, some nodes are designated as \emph{query} nodes~$q$.
Its (current) non-neighbors $v$ are sorted by decreasing~$s(q,v)$.
We are primarily interested in LP systems that can retrieve a small number of $K$ nodes with largest $s(q,v)$ for all $q$ in $o(N^2)$ time.

\section{Proposed Approach}
\label{sec:PermGNN}

In this section, we first state the limitations of GNNs. Then, we present our method for obtaining high quality node embeddings, with better representational power than GNNs.

\subsection{GNNs and Their Limitations}

GNNs start with a graph and per-node features $\fb_u$ to obtain a neighborhood-sensitive node representation $\xb_u$ for $u\in\Vcal$.  To meaningfully compare $\xb_u$ and $\xb_v$ and compute $s(u,v)$, information from neighbors of $u$ (and $v$) should be aggregated in such a way that the embeddings become invariant to permutations of the neighbors of $u$ (and $v$).
GNNs ensure permutation invariance by additive aggregation.
Given an integer $K$, for each node $u$, a GNN aggregates structural information $k$ hops away from $u$ to cast it into~$\xb_u$ for $k\le K$. Formally, a GNN first computes intermediate embeddings
$\set{\zb_u(k)\,|\, k\in [K]}$  in an iterative manner and then computes $\xb_u$,  using the following recurrent propagation rule. 
\begin{align}
\overline{\zb}_u(k-1)  &= \aggregate\big(\set{\zb_v(k-1) \,|\, v\in\nbr(u)}\big);\label{eq:gnn1}\\[-0.1ex]
\zb_u(k) &= \combine_1\big(\zb_u(k-1),\; \overline{\zb}_u(k-1)  \big);\label{eq:gnn3}\\[-0.1ex]
\xb_u &= \combine_2(\zb_u(1),\ldots,\zb_u(K))\label{eq:gnn4}
\end{align}
Here, for each node $u$ with feature vector $\fb_u$, we initialize $\zb_u(0) = \fb_u$; $\aggregate$ and $\combine_{1,2}$ are neural networks.  To ensure permutation invariance of the final embedding $\xb_u$, $\aggregate$ aggregates the intermediate $(k-1)$-hop information $\zb_v(k-1)$ with an additive (commutative, associative) function, guided by set function principles~\citep{zaheer2017deep}:
\begin{multline}
\aggregate\big(\set{\zb_v(k-1) \,|\, v\in\nbr(u)}\big) \\[-1ex]
= \sigma_1\left(\textstyle 
\sum_{v\in\nbr(u)} \sigma_2\big(\zb_v(k-1)\big) \right).
\label{eq:agg}
\end{multline}
Here $\sigma_1,\sigma_2$ are nonlinear activations.  In theory 
\citep[Theorem~2]{zaheer2017deep}, if $\combine_{1,2}$ are given `sufficient' hidden units, this set representation is universal.
In practice, however, commutative-associative aggregation suffers from limited expressiveness \citep{PabbarajuJ2019permute,wagstaff2019limitations,garg2020generalization,cohenkarlik2020regularizing}, which degrades the quality of $\xb_u$ and~$s(\cdot,\cdot)$, as described below.
Specifically, their expressiveness is constrained from two perspectives.

\paragraph{Attenuation of important network signals} 
GNNs are established to be intrinsically low pass filters~\cite{nt2019revisiting, wu2019simplifying}.  Consequently, they can attenuate high frequency signals which may contain crucial
structural information about the network.  To illustrate, assume that the node $u$ in Eqs.~\eqref{eq:gnn1}--\eqref{eq:gnn4} has two neighbors $v$ and $w$ and $\zb_v(k-1)=[+1, -1]$ and $\zb_w(k-1)=[-1, +1]$, which induce high frequency signals around the neighborhood of $u$. In practice, these two representations may carry important signals about the network structure. However, popular choices of $\sigma_2$
often diminish the effect of each of these vectors. In fact, the widely used linear form of $\sigma_2$ \citep{hamilton2017inductive, kipf2016semi} would completely annul their effects (since $\sigma_2(\zb_v(k-1)) + \sigma_2(\zb_w(k-1))=\bm{0}$) in the final embedding~$\xb_u$, which would consequently lose capacity for encapsulating neighborhood information.

\paragraph{Inability to distinguish between correlation structures}
In Eq.~\eqref{eq:agg}, the outer nonlinearity $\sigma_1$ operates over the sum of all representations of neighbors of~$u$.  Therefore, it cannot explicitly model the variations between the joint dependence of these neighbors.  Suppose the correlation between $\zb_{v}(k-1)$ and $\zb_{w}(k-1)$ is different from that between $\zb_{v'}(k-1)$ and $\zb_{w'}(k-1)$ for $\set{v,v',w,w'} \subseteq \nbr(u)$.  The additive aggregator in Eq.~\eqref{eq:agg} cannot capture the distinction.

Here, we develop a mitigation approach which exploits sequential memory, e.g., LSTMs, even though they are order-sensitive, and then neutralize the order sensitivity by presenting adversarial neighbor orders. 
An alternative mitigation approach is to increase the capacity of the aggregator (while keeping it order invariant by design) by explicitly modeling dependencies between neighbors, as has been attempted in image or text applications \cite{lee2019set, bloem2019probabilistic, shi2020deep, stelznergenerative}.


\subsection{Our Model: {\protect\our{}}}
Responding to the above limitations of popular GNN models, we design \our, the proposed adversarial permutation guided node embeddings.
\subsubsection*{Overview.} 
\begin{figure}
\centering\resizebox{0.48\textwidth}{!}{
\begin{tikzpicture}[>=latex]
  \def\nbrhd#1#2{%
    \begin{tikzpicture}
      \node [circle,fill=gray!20,inner sep=0, outer sep=0] (#1) {$#2$};
      \node [inner sep=0, outer sep=0, right=2mm of #1] (fe#1) {$\vdots$};
      \node [inner sep=0, outer sep=0, above=0mm of fe#1] (fne#1) {$\bm{f}$};
      \node [inner sep=0, outer sep=0, below=0mm of fe#1] (fse#1) {$\bm{f}$};
      \draw (#1) to (fe#1); \draw (#1) to (fne#1); \draw (#1) to (fse#1);
      \draw [decorate,decoration={brace,amplitude=4pt}]
      (fne#1.north east) -- (fse#1.south east)
      node (box#1) [midway,xshift=3pt] {};
      \node [anchor=center, right=2mm of box#1.east] (ffv#1) {$\bm{F}_{#2}$};
      \draw (box#1.east) -- (ffv#1.west);
      \node [anchor=center, fill=red!10, draw=red!40,
        inner sep=2pt, right=3mm of ffv#1] (Pphi#1) {$T_\phi$};
      \draw [->] (ffv#1.east) -- (Pphi#1.west);
      \node [anchor=center, draw=green!50, fill=green!10,
        inner sep=2pt, right=3mm of Pphi#1] (lstm#1) {$\text{LSTM}_\theta$};
      \draw [->] (Pphi#1.east) -- (lstm#1);
      \node [anchor=center, right=4mm of lstm#1, draw=green!50,
      fill=green!10] (sigma#1) {$\sigma_\theta$};
      \draw [->] (lstm#1.east) -- (sigma#1.west);
      \node [anchor=center, right=4mm of sigma#1] (xb#1) {$\bm{x}_{#2}$};
      \draw [->] (sigma#1.east) -- (xb#1.west);
  \end{tikzpicture}}
  
  \node [inner sep=0,outer sep=0, fill=gray!8]
  (vplus) {\nbrhd{vpluss}{v_+}};
  \node [inner sep=0,outer sep=0, fill=gray!8,
    below=1mm of vplus.south east, anchor=north east]
  (u) {\nbrhd{us}{u}};
  \node [inner sep=0,outer sep=0, fill=gray!8,
    below=1mm of u.south east, anchor=north east]
  (vminus) {\nbrhd{vminuss}{v_-}};
  \draw [thick] (vplus.west) -- (u.west)
  node [midway,xshift=-4mm] (edge) {edge};
  \draw [thick,dashed] (vminus.west) -- (u.west)
  node [midway,xshift=-7mm] (nonedge) {non-edge};  
  \node [above right=2mm and 2mm of u, outer sep=0,
  inner sep=0] (uvplus) {$\odot$};
  \draw [->] (vplus.east) -- (uvplus);
  \draw [->] (u.east) -- (uvplus);
  \node [below right=2mm and 2mm of u, outer sep=0,
  inner sep=0] (uvminus) {$\odot$};
  \draw [->] (vminus.east) -- (uvminus);
  \draw [->] (u.east) -- (uvminus);
  \node [anchor=center, inner sep=0pt, outer sep=0pt,
        right=2mm of uvplus] (simuvplus) {$\text{sim}(u,v_+)$};
  \draw [->] (uvplus.east) -- (simuvplus.west);
  \node [anchor=center, inner sep=0pt, outer sep=0pt,
        right=2mm of uvminus] (simuvminus) {$\text{sim}(u,v_-)$};
  \draw [->] (uvminus.east) -- (simuvminus.west);
  \node [anchor=center, draw, inner sep=2pt, outer sep=0pt,
        right=28mm of u.east] (relu) {ReLU};
  \draw [->] (simuvplus.south east) -- (relu.north west);
  \draw [->] (simuvminus.north east) -- (relu.south west);
  \node [anchor=center, inner sep=0pt, outer sep=0pt,
    right=3mm of relu.east] (loss) {loss};
  \draw [->] (relu.east) -- (loss.west);
  \node [left=2mm of relu.west, inner sep=0,
  outer sep=0] (margin) {$\Delta$};
  \draw [->] (margin.east) -- (relu.west);
  \node [right=4mm of u.east, fill=blue!10, draw=blue!40] (Cpsi) {$C_\psi$};
  \draw [->] (u) -- (Cpsi);
  \node [right=3mm of Cpsi] (beeu) {$\bm{b}_u$};
  \draw [->] (Cpsi) -- (beeu);
\end{tikzpicture}
}
\caption{\our{} min-max loss and hashing schematic.}
\label{fig:PermGnnLossSchematic}
\end{figure}
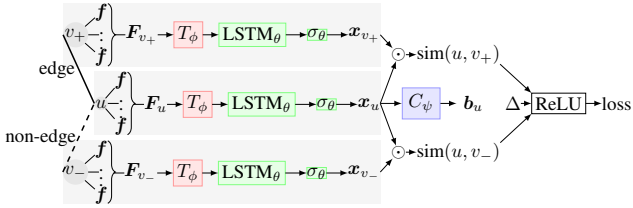

\hspace{-2mm}
Given a node $u$, we first compute an embedding $\xb_u$ using a \emph{sequence} encoder, parameterized by $\theta$:
\begin{align}
\xb_u = \rho_{\theta}\big(\set{\fb_v \,|\, v\in\nbr(u)}\big),
\end{align}
where $\nbr(u)$ is presented in some arbitrary order (to be discussed).  In contrast to the additive aggregator, $\rho$ is modeled by an LSTM \citep{HochreiterS1997LSTM}, followed by a fully-connected feedforward neural network (See Figure~\ref{fig:PermGnnLossSchematic}).  Such a formulation captures the presence of high frequency signal in the neighborhood of $u$ and the complex dependencies between the neighbors $\nbr(u)$ by combining their influence via the recurrent states of the LSTM.

However, now the embedding $\xb_u$ is no longer invariant to the permutation of the neighbors $\nbr(u)$.  As we shall see, we counter this by casting the LP objective as an instance of a min-max optimization problem. Such an adversarial setup refines $\xb_u$ in an iterative manner, to ensure that the resulting trained embeddings are permutation invariant (at least as far as possible in a non-convex optimization setting).

\subsubsection*{\our{} architecture.}
Let us suppose $\pib = [\pi_1,...,\pi_{|\nbr(u)|}] \in \Pi_{|\nbr(u)|}$ is some arbitrary permutation of the neighbors of node~$u$.  We take the features of neighbors of $u$ in the order specified by $\pib$, i.e., $\bigl(v_{\pi_1}, v_{\pi_2}, \ldots, v_{\pi_{|\nbr(u)|}}\bigr)$, and pass them into an LSTM: 
\begin{align}
\hspace{-2mm}   \yb _{u,1},..., \yb _{u,{|\nbr(u)|}} = \textsc{LSTM}_{\theta}\big(\fb_{v_{\pi_1}}, ..., \fb_{v_{\pi_{|\nbr(u)|} }}\big). \label{eq:lstm}
\end{align}
Here $\big(\yb _{u, k} \big)_{k\in [|\nbr(u)|]}$ is a sequence of intermediate representation of node $u$, which depends on the permutation~$\pib$.  Such an approach ameliorates the limitations of GNNs in two ways:\\
(1) Unlike GNNs, the construction of $\yb_{\bullet}$ is not limited to symmetric aggregation, and is therefore able to capture crucial network signals including those with high frequency~\cite{borovkova2019ensemble}. \\
(2) An LSTM (indeed, any RNN variant) is designed to capture the influence of one token of the sequence on the subsequent tokens. In the current context, the state variable $\hb_{k}$ of the LSTM combines the influence of first $k-1$ neighbors in the input sequence, \ie, $v_{\pi_1},\ldots v_{\pi_{k-1}}$ on the $k$-th neighbor $v_{\pi_k}$. Therefore, these recurrent states allow  $\yb_{\bullet}$ to capture the complex dependence between the features $\fb_\bullet$.\\
%
Next, we compute the final embeddings $\xb_u$ by using an additional nonlinearity on the top of the sequence $(\yb_{u, k} )_{k\in [|\nbr(u)|]}$ output by the LSTM:
\begin{align}
\xb_{u;\pib} = \sigma_{\theta} \big(\yb_{u, 1}, \yb_{u, 2}, \ldots, \yb_{u, |\nbr(u)|}\big) \in \mathbb{R}^D. \label{eq:sigma-theta-intro}
\end{align}
Note that the embeddings $\set{\xb_u}$ computed above depends on $\pib$, the permutation of the neighbors $\nbr(u)$ given as the input sequence to the LSTM in Eq.~\eqref{eq:lstm}. 

\paragraph{Removing the sensitivity to~$\pib$}
One simple way to ensure permutation invariance is to compute the average of $\xb_{u;\pib}$ over all permutations~$\pib \in \Pi_{|\nbr(u)|}$, similar to~\citet{murphy2019janossy}.  At a time and space complexity of at least $O(\sum_{u\in\Vcal} |\Pi_{|\nbr(u)|}|)$, this is quite impractical for even moderate degree nodes.  Replacing the exhaustive average by a Monte Carlo sample does improve representation quality, but is still very expensive.
{\citet{murphy2019relational} proposed a method called $\pi$-SGD, which samples one permutation per epoch.  While it is more efficient than sampling multiple permutations, it shows worse robustness in practice.}

\subsubsection*{Adversarial permutation-driven LP objective.}
Instead of brute-force sampling, we setup a two-party game, one being the network for LP, vulnerable to $\pib$, and {the other} being an adversary, which tries to make the LP network perform poorly by choosing a `bad'~$\pib$ at each node.
\begin{tcolorbox}[colframe=gray!40,boxsep=0mm,left=0pt,right=0pt,top=0pt,bottom=0pt]
\begin{algorithmic}[1]
\State pick initial $\pib^u$ at each node~$u$
\Repeat
\State fix $\set{\pib^u: u\in\Vcal}$;
optimize $\theta$ for best LP accuracy
\State fix $\theta$; find next $\pib^u$ at all $u$ for worst LP accuracy
\Until{LP performance stabilizes}
\end{algorithmic}
\end{tcolorbox}
\noindent
Let $\pib^u \in \Pi_{|\nbr(u)|}$ be the permutation used to shuffle the neighbors of $u$ in Eq.~\eqref{eq:lstm}.  Conditioned on $\pib^u, \pib^v$, we compute the score for a node-pair $(u,v)$ as 
\begin{align}
s_{\theta}(u,v|\pib^u,\pib^v)
&=\similarity(\xb_{u;\pib^u}, \xb_{v;\pib^v}),  \label{eq:SimCos}
\end{align}
where $\similarity(\ab,\bb)$ denotes the cosine similarity between $\ab$ and~$\bb$.  To train our LP model to give high quality ranking, we consider the following AUC loss surrogate \cite{Joachims2005multivariate}: 
\begin{align}
& \loss(\theta; \set{\pi^{w}}_{w\in\Vcal} )  \nn  \\
&= \!\!\! \sum_{  \substack{(u,v)\in \Ecal\\ (r,t)\in\ne}}
\Big[ \Delta  + s_{\theta}(r,t)|\pib^r,\pib^t  
  - s_{\theta}(u,v|\pib^u,\pib^v) \Big]_{+}
\label{eq:HingeRankingLoss}
\end{align}
where $\Delta$ is a tunable margin and $[a]_+=\max\set{0,a}$.

As stated above, we aim to train LP model parameters $\theta$ in such a way that the trained embeddings $\set{\xb_u}$ become invariant to the permutations of $\nbr(u)$ for all nodes $u\in\Vcal$.  This requirement suggests the following min-max loss:
\begin{align}
\min_{\theta}    \max_{\set{\pib^{w}}_{w\in\Vcal}} \loss(\theta; \set{\pib^{w}}_{w\in\Vcal}). \label{eq:MinMaxOptHard}
\end{align}

\subsubsection*{Neural permutation surrogate.}
As stated, the complexity of Eq.~\eqref{eq:MinMaxOptHard} seems no better than exhaustive enumeration of permutations.  To get past this apparent blocker, just as max is approximated by softmax (a multinomial distribution), a `hard' permutation (1:1 assignment) $\pib^w$ is approximated by a `soft' permutation matrix $\Pb^w$ --- a doubly stochastic matrix --- which allows continuous optimization. 

Suppose $\Fb_w =\big[\fb_{v_1},\fb_{v_2},\ldots,\fb_{v_{|\nbr(w)|}}\big]$ is a matrix whose rows are formed by the features of $\nbr(w)$ presented in some canonical order. Then $\Pb^w \Fb_w$ approximates a permuted feature matrix corresponding to some permuted sequence of neighbor feature vectors.  The RHS of Eq.~\eqref{eq:lstm} can be written as $\text{LSTM}_\theta(\Pb^w \Fb_w)$, which eventually lets us express loss as a function of~$\Pb^w$.  We can thus rewrite the min-max optimization \eqref{eq:MinMaxOptHard}~as 
\begin{align}
\min_{\theta}
\max_{\set{\Pb^w \,|\, w\in\Vcal}} \loss(\theta; \set{\Pb^{w}}_{w\in\Vcal}), \label{eq:MinMaxOptSoft}
\end{align}
where the inner maximization is carried out over all `soft' permutation matrices $\Pb^w$, parameterized as follows.

In deep network design, a trainable multinomial distribution is readily obtained by applying a softmax to trainable (unconstrained) logits.  Analogously, a trainable soft permutation matrix $\Pb^w$ can be obtained by applying a  Gumbel-Sinkhorn network `GS' \citep{Mena+2018GumbelSinkhorn} to a trainable (unconstrained) `seed' square matrix, say,~$\Ab^w$:
\begin{align}
&\Pb^w = \lim_{n\to\infty} \text{GS}^n(\Ab^w), \quad \text{where} \nn \\
&\text{GS}^0(\Ab^w) = \exp(\Ab^w) \quad \text{and} \nn \\
&\text{GS}^n(\Ab^w) = \ColScale\left(
\RowScale\big(\text{GS}^{n-1}(\Ab^w)\big) \right). \nn
\end{align}
Here, $\ColScale$ and $\RowScale$ represent column and row normalization. $\text{GS}^n(\Ab^w)$ is the doubly stochastic matrix obtained by consecutive row and column normalizations of~$\Ab^w$.  It can be shown that
\begin{align}
\lim_{n\to\infty} \text{GS}^n(\Ab^w) &=
\argmax_{\Pb \in \Pcal_{|\nbr(w)|}}\text{Tr}\left[\Pb^\top \Ab^w\right].
\end{align}
$\text{GS}^n$ thus represents a recursive differentiable operator that permits backpropagation of $\loss$ to $\set{\Ab^w}$.  In practice, $n$ is a finite hyperparameter, the larger it is, the closer the output to a `hard' permutation.

Allocating a separate unconstrained seed matrix $\Ab^w$ for each node $w$ would lead to an impractically large number of parameters.  Therefore, we express $\Ab^w$ using a globally shared network $\Tb_\phi$ with model weights $\phi$, and the per-node feature matrix $\Fb_w$ already available.  I.e., we define 
\begin{align}
\Ab^w := \Tb_\phi(\Fb_w / \tau), \label{eq:TIntro}
\end{align}
where $\tau>0$ is a temperature hyperparameter that encourages $\text{GS}^n(\Ab^w)$ toward a `harder' soft permutation.  The above steps allow us to rewrite optimization \eqref{eq:MinMaxOptSoft} in terms of $\theta$ and $\phi$ in the form $\min_\theta \max_\phi \loss(\theta; \phi)$.  After completing the min-max optimization, the embedding $\xb_u$ of a node $u$ can be computed using some arbitrary neighbor permutation.  By design, the impact on $\similarity(u,v)$ is small when different permutations are used.

\section{Scalable LP by Hashing Representations}
\label{sec:LSH}
At this point, we have obtained representations $\xb_u$ for each node~$u$ using \our. Our next goal is to infer some number of most likely future edges.

\paragraph{Prediction using exhaustive comparisons} Here, we first enumerate the scores for all possible potential edges (the current non-edges) and then report top-$K$ neighbors for each node. Since most real-life social networks are sparse, potential edges can be $\Theta(|\Vcal|^2)$ in number.  Scoring all of them in large graphs is impractical; we must limit the number of comparisons between potentially connecting node pairs to be as small as possible. 
%
\subsection{Data-Oblivious LSH with Random Hyperplanes}
\label{sec:Hyperplanes}
When for two nodes $u$ and $v$, $\similarity(u,v)$ is defined as $\cos(\xb_u, \xb_v)$ with $\xb_\bullet\in\RR^D$, the classic random hyperplane LSH can be used to hash the embeddings $\xb_\bullet$. Specifically, we first draw  $H$ uniformly random hyperplanes passing through the origin in the form of their unit normal vectors $\bm{n}_h \in \mathbb{R}^D, h\in[H]$~\cite{Charikar2002lsh}. Then we set $b_u[h] = \sign(\bm{n}_h \cdot \xb_u) \in \pm1$ as a 1-bit hash and $\bm{b}_u \in \pm1^H$ as the $H$-bit hash code of node~$u$. Correspondingly, we set up $2^H$ hash buckets with each node going into one bucket.  If the buckets are balanced, we expect each to have $N/2^H$ nodes.  Now we limit pairwise comparisons to only node pairs within each bucket, which takes $N^2/2^H$ pair comparisons.  By letting $H$ grow slowly with $N$, we can thus achieve sub-quadratic time. 
However, such a hashing method is data oblivious--- the hash codes are not learned from the distribution of the original embeddings $\xb_\bullet$. It performs best when the embeddings are uniformly dispersed in the $D$-dimensional space, so that the random hyperplanes can evenly distribute the nodes among several hash buckets.

\subsection{Learning Data-Sensitive Hash Codes}
\label{sec:HashOpt}
To overcome the above limitation of random hyperplane based hashing, we devise a data-driven learning of hash codes as explored in other applications~\citep{WeissTF2009SpectralHashing}.  
Specifically, we aim to design an additional transformation of the vectors $\set{\xb_u}$ into compressed representations $\set{\bm{b}_u}$, with the aim of better balance across hash buckets and reduced prediction time.

\subsubsection*{Hashing/compression network.}
In what follows, we will call the compression network $C_\psi: \mathbb{R}^D\to [-1,1]^H$, with model parameters~$\psi$.  We interpret $\sign\big(C_\psi(\xb_u)\big)$ as the required binary hash code $\bb_u\in\set{-1,+1}^H$, with the surrogate $\tanh(C_\psi(\xb_u))$, to be used in the following smooth optimization:
\begin{multline}
\hspace{-2mm}\min_\psi \textstyle
\frac{\alpha}{|\Vcal|}
\sum_{u\in \Vcal} \big| \bm{1}^\top 
\tanh(C_\psi(\xb_u)) \big| \\[-1ex]
+ \textstyle \frac{\beta}{|\Vcal|}
\sum_{u\in\Vcal} \Big\|
\big| \tanh(C_\psi(\xb_u)) \big| - \bm{1} \Big\|_1 \\
+ \textstyle \frac{\gamma}{|\overline{E}|}
\sum_{(u,v) \in \overline{E}} \left|\tanh(C_\psi(\xb_u)) \cdot
\tanh(C_\psi(\xb_v))  \right|
\label{eq:HashOpt}
\end{multline}
Here, $\overline{E}$ is the set of non-edges and $\alpha, \beta, \gamma \in (0,1)$, with $\alpha+\beta+\gamma=1$ are tuned hyperparameters. 
%
The final binary hash code $\bb_u=\text{sign}(C_\psi(\xb_u))$.  The salient terms in the objective above seek the following goals.

\noindent{\bfseries \slshape Bit balance:} If each bit position has as many $-1$s as $+1$s, that bit evenly splits the nodes.  The term $\big| \bm{1}^\top \tanh(C_\psi(\xb_u)) \big|$ tries to bit-balance the hash codes.

\noindent{\bfseries \slshape No sitting on the fence:} The optimizer is prevented from setting $\bm{b}=\bm{0}$ (the easiest way to balance it) by including a term $\sum_h \big| |\bm{b}[h]| - 1 \big| = \big\| |\bm{b}| - \bm{1} \big\|_1$. 

\noindent{\bfseries \slshape Weak supervision:} 
The third term encourages currently unconnected nodes to be assigned dissimilar bit vectors.



\subsubsection*{Bucketing and ranking.}
Note that, we do not expect the dot product between the learned hash codes $\bm{b}_u \cdot \bm{b}_v$ to be a good approximation for $\cos(\xb_u,\xb_v)$, merely that node pairs with large  $\cos(\xb_u,\xb_v)$ will be found in the same hash buckets.  We form the buckets using the recipe of \citet{GionisIM1999hash}.  We adopt the high-recall policy that node-pair $u,v$ should be scored if $u$ and $v$ share at least one bucket.  Algorithm~\ref{algo:LshTopK} shows how the buckets are traversed to generate and score node pairs, then placed in a heap for retrieving top-$K$ pairs.
Details can be found in the Appendix.

\begin{algofig}
\begin{tcolorbox}[colframe=gray!40,boxsep=0mm] \footnotesize
\small
\begin{algorithmic}[1]
\State \textbf{Input}: Graph $\Gcal=(\Vcal, \Ecal)$;
binary hash-codes $\set{\bb_u}$;
query nodes $\Qcal$; the number ($K$) of nodes to be recommended per query node
\State \textbf{Output}: Ranked recommendation list $R_q$ for all $q\!\in\!\Qcal$
\gap
\State initialize LSH buckets
\For{$u\in\Vcal$}
\State add $u$ to appropriate hash buckets
\EndFor

\For{$q\in\Qcal$}
\State initialize score heap $H_q$ with capacity~$K$
\EndFor

\For{each LSH bucket $B$}
\For{$(u,v)\in B $}
\If{$u\in\Qcal$}
\State insert $\langle v, s(u,v)\rangle$ in $H_u$;
prune if $|H_u|\!>\!K$
\EndIf
\If{$v\in\Qcal$}
\State insert $\langle u, s(u,v)\rangle$ in $H_v$;
prune if $|H_v|\!>\!K$
\EndIf
\EndFor
\EndFor
\gap
\For{$q\in\Qcal$}
 \State sort $H_q$ by decreasing score to get ranked list~$R_q$
\EndFor
\gap
\State \textbf{return} $\set{R_q |q \in\Qcal}$
\end{algorithmic}
\end{tcolorbox}
\caption{Reporting ranked list of potential edges fast.}
\label{algo:LshTopK}
\end{algofig}


\begin{table*}[t]
\begin{center}
\maxsizebox{0.81\hsize}{!}{
\begin{tabular}{l|ccccc|ccccc}
\hline
& \multicolumn{5}{c|}{\textbf{Mean Average Precision (MAP)}} & \multicolumn{5}{c}{\textbf{Mean Reciprocal Rank (MRR)}} \\

&Twitter  &  \google & Cora & Citeseer  & PB 
&Twitter  &  \google & Cora & Citeseer  & PB  \\ \hline\hline
AA & 0.727 & 0.321 & 0.457 & 0.477 & \best{0.252} & 0.904 & {0.553} & \best{0.535} & 0.548 & 0.508 \\
CN & 0.707 & 0.292 & 0.377 & 0.401 & 0.218 & \best{0.911} & 0.553 & 0.460 & 0.462 & \best{0.516} \\\hline
Node2Vec & 0.673 & 0.330 & 0.448 & 0.504 & 0.182 & 0.832 & 0.551 & 0.484 & 0.546 & 0.333 \\
DeepWalk & 0.624 & 0.288 & 0.432 & 0.458 & 0.169 & 0.757 & 0.482 & 0.468 & 0.492 & 0.303 \\\hline
GraphSAGE & 0.488 & 0.125 & 0.393 & 0.486 & 0.077 & 0.638 & 0.233 & 0.425 &  0.523 & 0.156 \\
GCN & 0.615 & 0.330 & 0.408 & 0.464 & 0.200 & 0.789 & 0.482 & 0.444 & 0.505 & 0.345 \\
Gravity & \best{0.735} & 0.360 & 0.407 & 0.462 & 0.193 & 0.881 & 0.540 & 0.438 & 0.518 & 0.330 \\\hline
\our & \best{0.735} & \best{0.385} & \best{0.480} & \best{0.560} & 0.220 & 0.880 & \best{0.581} & 0.524 & \best{0.600} & 0.397 \\ \hline %
\end{tabular} }
\end{center}
\caption{MAP and MRR for all LP algorithms (\our and baselines) on the ranked list of all potential edges ($K=\infty$) across all five datasets, with 20\% test set.  Numbers in bold font indicate the best performer. }

\label{tab:main-map-mrr}
\end{table*}

\section{Experiments}
\label{sec:Expt}

We report on a comprehensive evaluation of \our{} and its accompanying hashing strategy. Specifically, we address the following research questions.
\begin{itemize*}
\item[\textbf{RQ1:}] How does the LP accuracy of \label{rq:prediction}\our{} compare with classic and recent link predictors? Where are the gains and losses?
\item[\textbf{RQ2:}] How \label{rq:PermGnnVsMultiPerm}does \our{} compare with brute-force sampling of neighbor permutations?
\item[\textbf{RQ3:}] Exactly \label{rq:permInvariance}where in our adversarially trained network is permutation insensitivity getting programmed?
\item[\textbf{RQ4:}] Does \label{rq:hashingBasics}the hashing optimization reduce prediction time, compared to exhaustive computation of pairwise scores?
\end{itemize*}

\subsection{Experimental Setup}
\subsubsection*{Datasets.}
We consider five real world datasets:
\begin{enumerate*}[label=(\arabic*)]
\item Twitter~\cite{leskovec2012learning}, 
\item \google~\cite{leskovec2010kronecker},
\item Cora~\cite{getoor2005link,sen2008collective},
\item Citeseer~\cite{getoor2005link,sen2008collective} and
\item PB~\cite{ackland2005mapping}.
\end{enumerate*}
%
\subsubsection*{Baselines.}
We compare \our{} with several hashable LP algorithms.
Adamic Adar (AA) and Common Neighbors (CN) \cite{LibenNowellK2007LinkPred} are classic unsupervised methods.
Node2Vec \cite{grover2016node2vec} and DeepWalk \cite{perozzi2014deepwalk} are node embedding methods based on random walks.
Graph Convolutional Network (GCN) \cite{kipf2016variational},
GraphSAGE \cite{hamilton2017inductive} 
Gravity~\cite{Salha+2019gravity} are node
embedding methods based on GNNs.
We highlight that SEAL~\citep{ZhangC2018LinkPredGNN}
does not readily lend itself to a hashable LP mechanism and therefore, we do not compare it in this paper.  


\subsubsection*{Evaluation protocol.}
%
Similar to the evaluation protocol of \citet{BackstromL2011SRW}, we partition the edge (and non-edge) sets into training, validation and test folds as follows.  For each dataset, we first build the set of query nodes $\Qcal$, where each query contains at least one triangle around it.
Then, for each $q\in\Qcal$, in the original graph, we partition the neighbors $\nbr(q)$ and the non-neighbors $\nnbr(q)$ which are  within 2-hop distance
from $q$ into 70\% training, 10\% validation and 20\% test sets, where the node pairs are sampled uniformly at random.  We disclose the resulting sampled graph induced by the training and validation sets to the LP model.  Then, for each query $q\in\Qcal$, the trained LP model outputs a top-$K$ list of potential neighbors from the test set. 
Using ground truth, we compute the average precision (AP) and reciprocal rank (RR) of each top-$K$ list. Then we average over all query nodes to get mean AP (MAP) and mean RR (MRR).

\subsection{Comparative Analysis of LP Accuracy}
First, we  address the research question \textbf{RQ1}
by comparing LP accuracy of \our against baselines, in terms of MAP and MRR across the datasets.

\subsubsection*{MAP and MRR summary.}
Table~\ref{tab:main-map-mrr} summarizes LP accuracy across all the methods. We make the following observations.
\begin{enumerate*}[label=(\arabic*)]
\item \our outperforms all the competitors in terms of MAP, in four datasets, except PB, where it is outperformed by AA. Moreover, in terms of MRR, it outperforms all the baselines for \google\ and Citeseer datasets.
\item The performance of GNNs are comparable for Cora and Citeseer.  Due to its weakly supervised training procedure, the overall performance of GraphSAGE is poor among the GNN based methods.
\item The classic unsupervised predictors, \ie, AA and CN often beat some recent embedding models. AA is the best performer in terms of MAP in PB and in terms of MRR in Twitter. Since AA and CN encourage triad completion, which is a key factor for growth of several real life networks, they often serve as good link predictors~\cite{SarkarCM2011LPembed}.
\item The random walk based embeddings, \emph{viz.} Node2Vec and DeepWalk, show moderate performance. Notably, Node2Vec is the second best performer in Citeseer. 
\end{enumerate*}

\begin{figure}[t]
\centering

\subfloat[\google]{
\includegraphics[width=.21\textwidth]{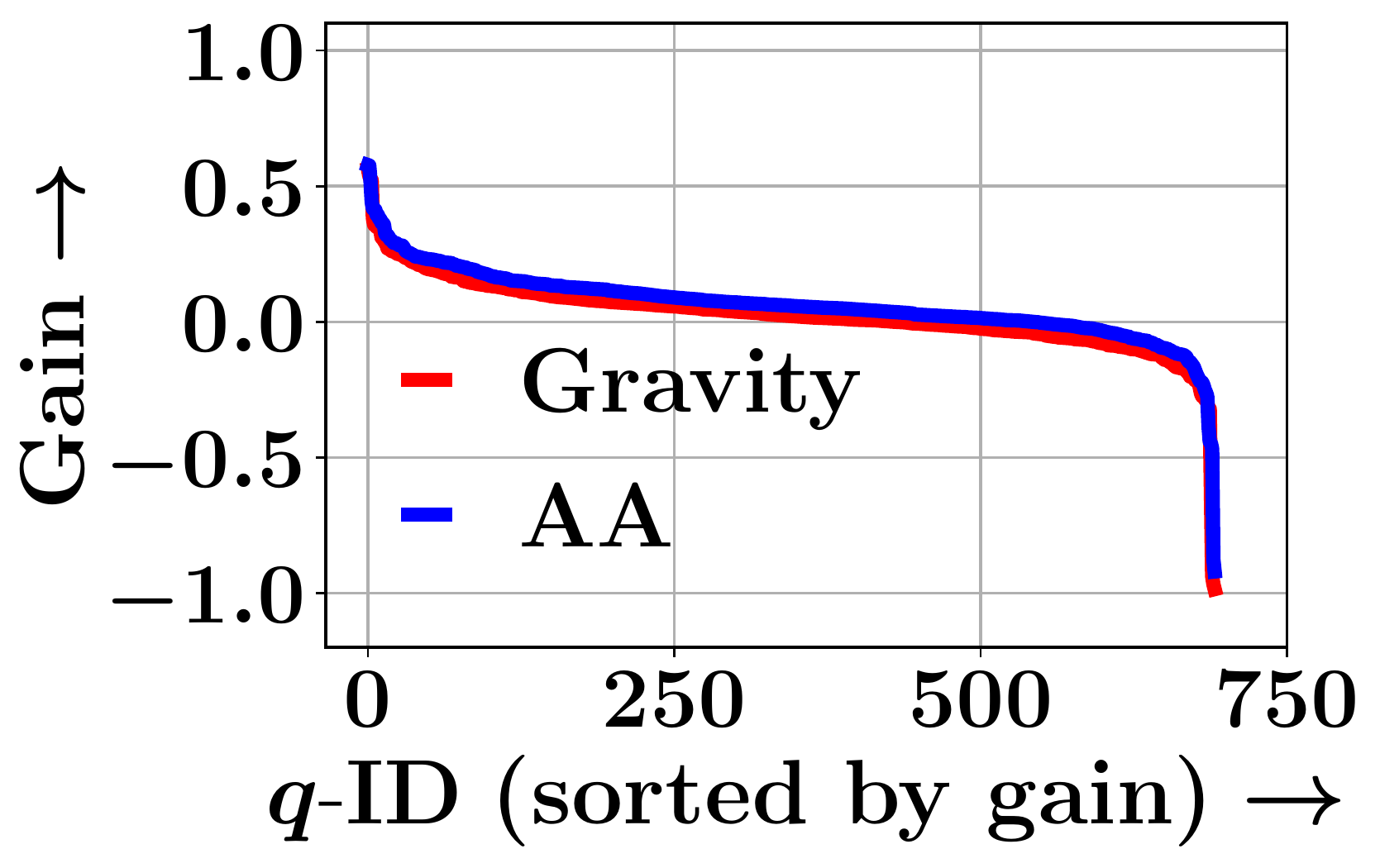}
}\hspace{2mm}
\subfloat[Citeseer]{
\includegraphics[width=.21\textwidth]{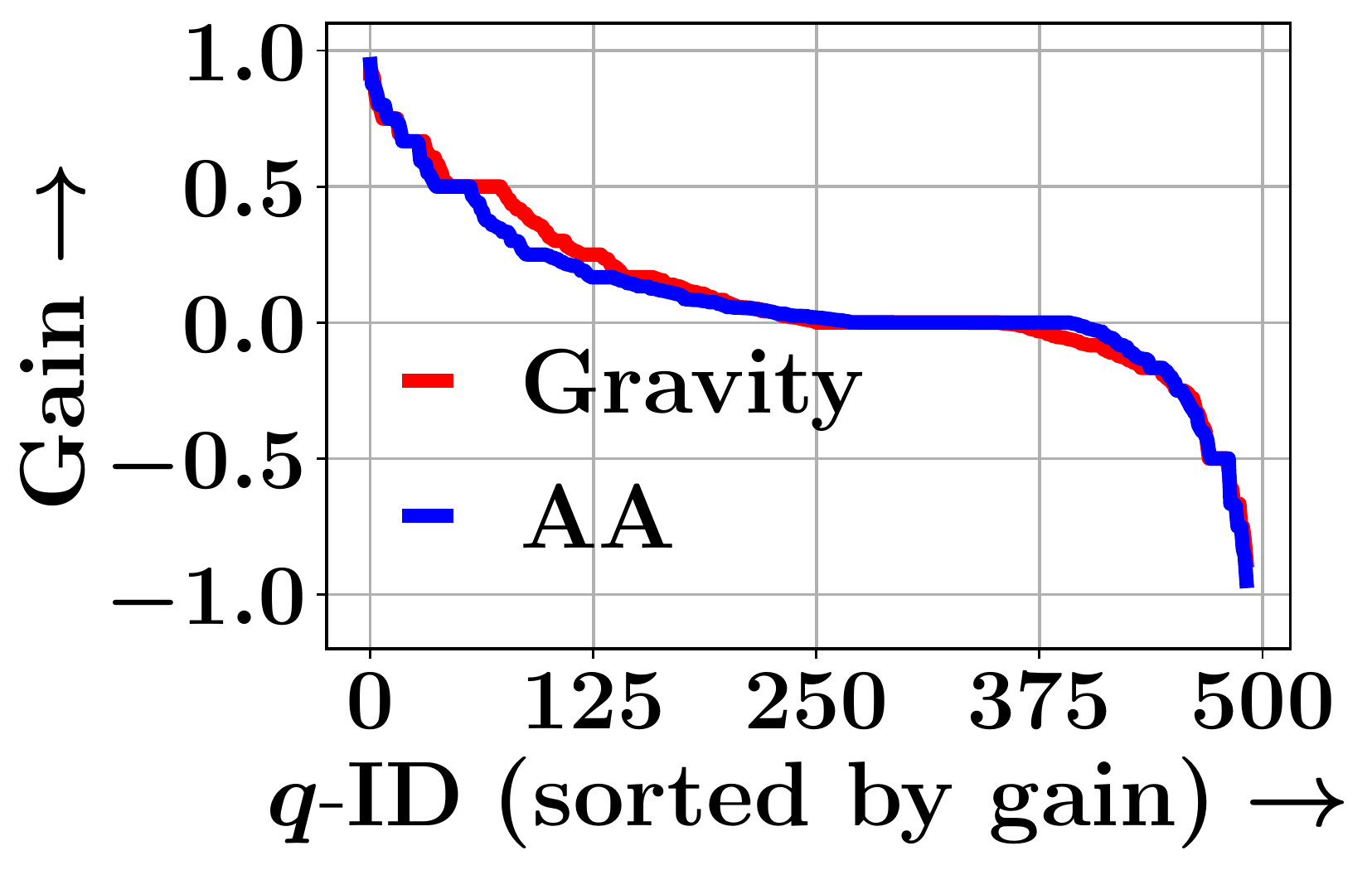}
}

\caption{Query-wise wins and losses in terms of $\text{AP}(\our)-\text{AP}(\text{baseline})$, the gain (above x-axis) or loss (below x-axis) of AP of \our with respect to competitive baselines. Queries $\Qcal$ are sorted by decreasing gain of \our{} along the $x$-axis. }
\label{fig:MapMrrDiff}
\end{figure}

\subsubsection*{Drill-down.}
Next, we compare ranking performance at individual query nodes. For each query (node) $q$, we measure the gain (or loss) of \our{} in terms of average precision, \ie, $\text{AP}(\our)-\text{AP}(\text{baseline})$ for three competitive baselines, across \google and  Citeseer datasets. From Figure~\ref{fig:MapMrrDiff}, we observe that, for \google and Citeseer respectively, \our\ matches or exceeds the baselines for 60\% and 70\% of the queries.


\begin{figure}[t]
\centering
\subfloat[Twitter]
{\includegraphics[width=.21\textwidth]{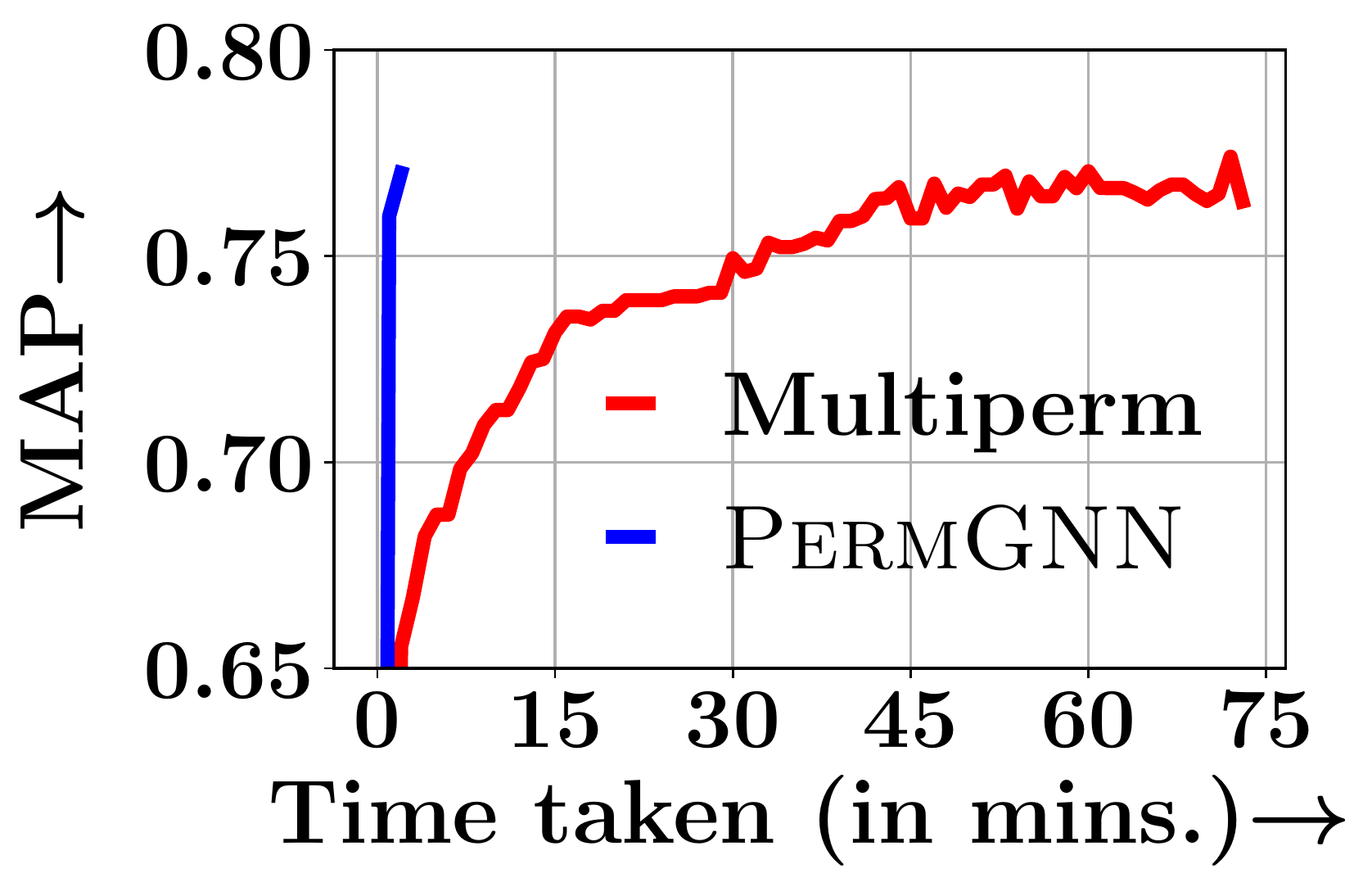}}\hspace{2mm}
\subfloat[\google]
{\includegraphics[width=.21\textwidth]{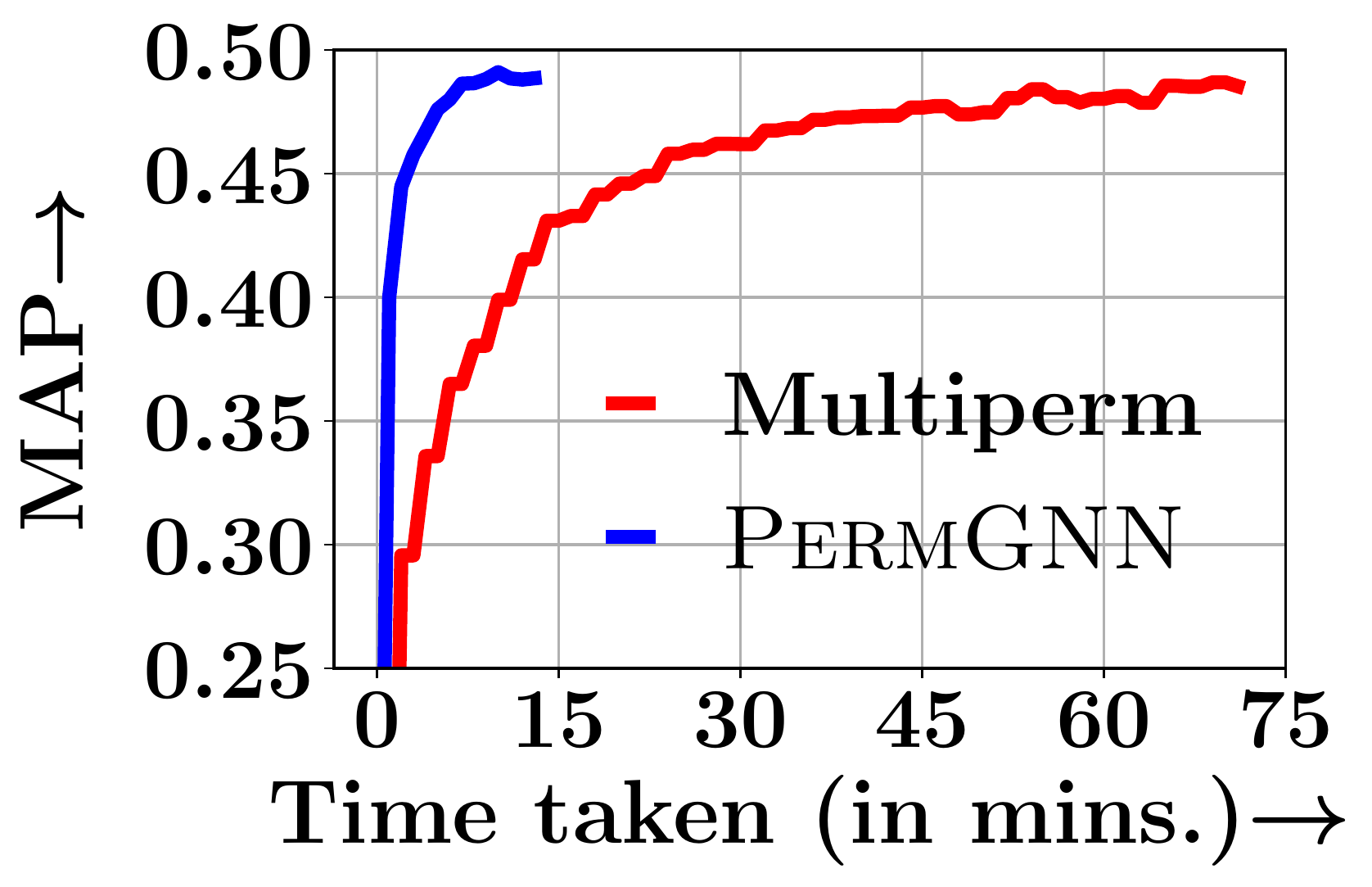}}

\caption{Validation MAP against training epochs for Twitter and \google.  \our converges faster than MultiPerm. 
}
\label{fig:PermGnnVsMultiPerm}
\end{figure}

\subsection{\our vs.\ Sampling Permutations}
Next, we address research question \textbf{RQ2}
by establishing the utility of \our{} against its natural alternative \textbf{MultiPerm}, in which a node embedding is computed by averaging permutation-sensitive representations over several sampled permutations.  Figure~\ref{fig:PermGnnVsMultiPerm} shows that \our is ${>}15{\times}$ and ${>}4.5{\times}$  faster than
the permutation averaging based method for Twitter and  \google datasets. MultiPerm also occupies significantly larger RAM than \our.
\begin{figure}[t]
\centering
\subfloat[Cora]{\includegraphics[width=.20\textwidth]{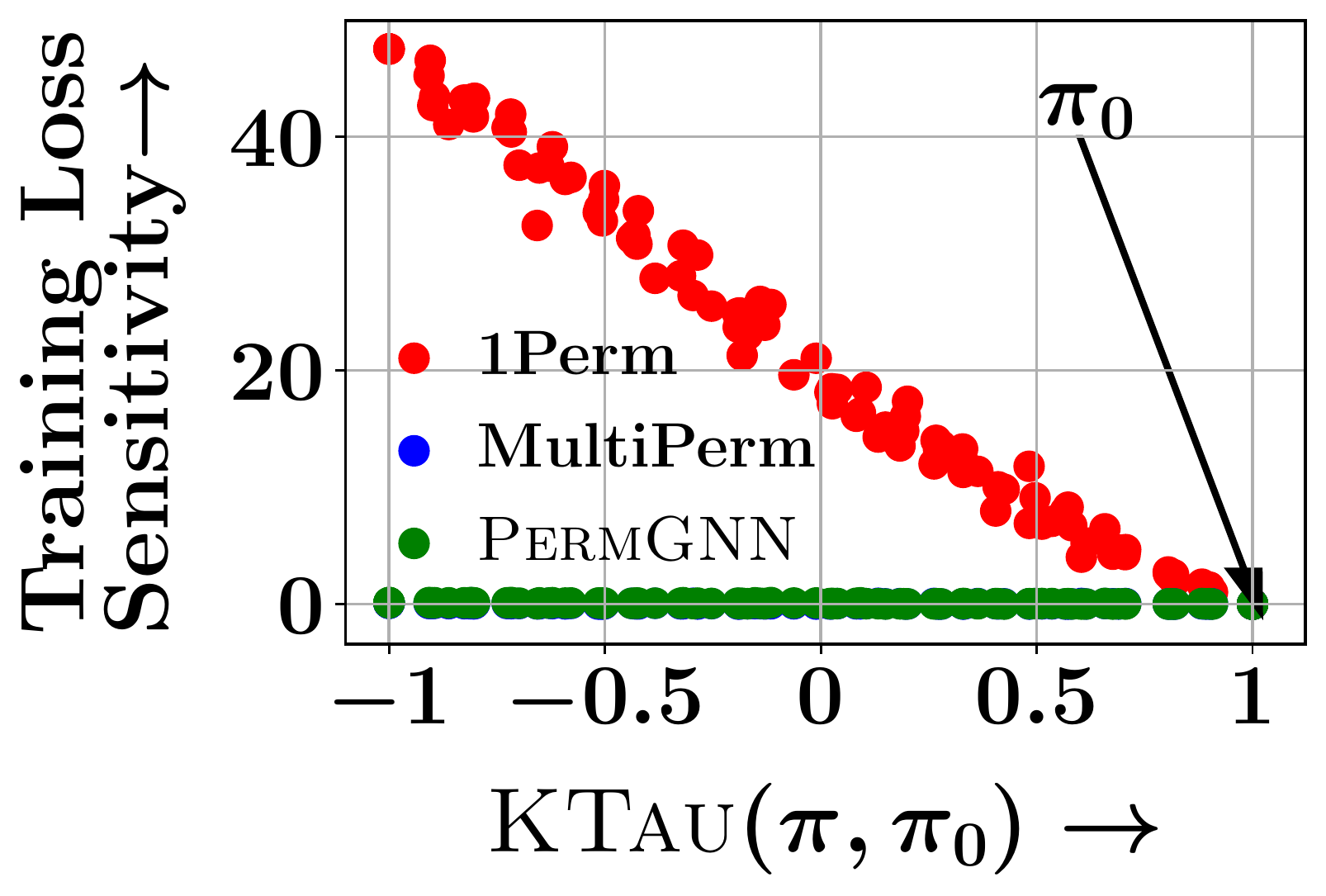}}\hspace{2mm}
\subfloat[Citeseer]{\includegraphics[width=.20\textwidth]{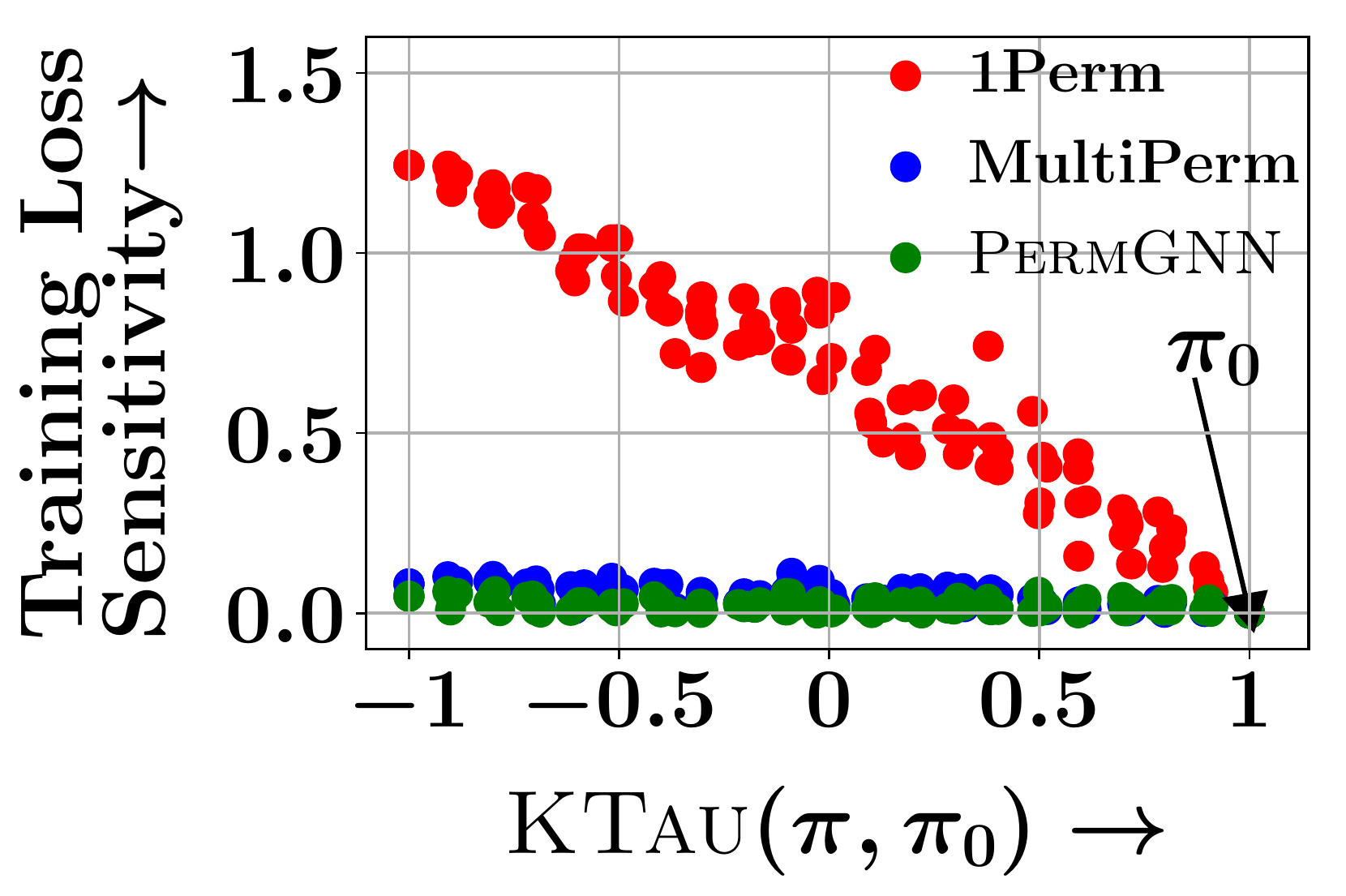}}

\caption{Effect of neighbor order perturbation on training loss.  As we move away from the canonical permutation $\pi_0$, training loss increases steeply for 1Perm, but remains roughly stable for MultiPerm and \our.}  
\label{fig:KtauVsMap}
\end{figure} 

\subsection{Permutation Invariance of \our}
Here, we answer the research question \textbf{RQ3}.
To that end, we first train \our along with its two immediate alternatives: (i)~\textbf{1Perm}, where a vanilla LSTM is trained with a single canonical permutation~$\pib_0$ of the nodes; and, (ii)~\textbf{Multiperm}, where an LSTM is trained using several sampled permutations of the nodes.  Then, given a different permutation $\pib$, we compute the node embedding $\xb_{u;\pib}$ by feeding the corresponding sequence of neighbors $\pib(\nbr(u))$ (sorted by node IDs of $\nbr(u)$ assigned by~$\pib$), as an input to the trained models.  Finally, we use these embeddings for LP and measure the relative change in training loss.  Figure~\ref{fig:KtauVsMap} shows a plot of  $(\textsc{loss}(\pib)-\textsc{loss}(\pib_0))/\textsc{loss}(\pib_0)$ against the correlation between $\pib$ and the canonical order $\pib_0$, measured in terms of Kendall's $\tau$, $\KTau(\pib,\pib_0)$. It reveals that 1Perm suffers a significant rise in training loss when the input node order $\pib$ substantially differs from the canonical order $\pib_0$, \ie, $\KTau(\pib,\pib_0)$ is low.  Both Multiperm and \our turns out to be permutation-insensitive across a wide range of node orderings.

\begin{figure}[t]
\centering
\subfloat[Cora]{\includegraphics[width=.20\textwidth]{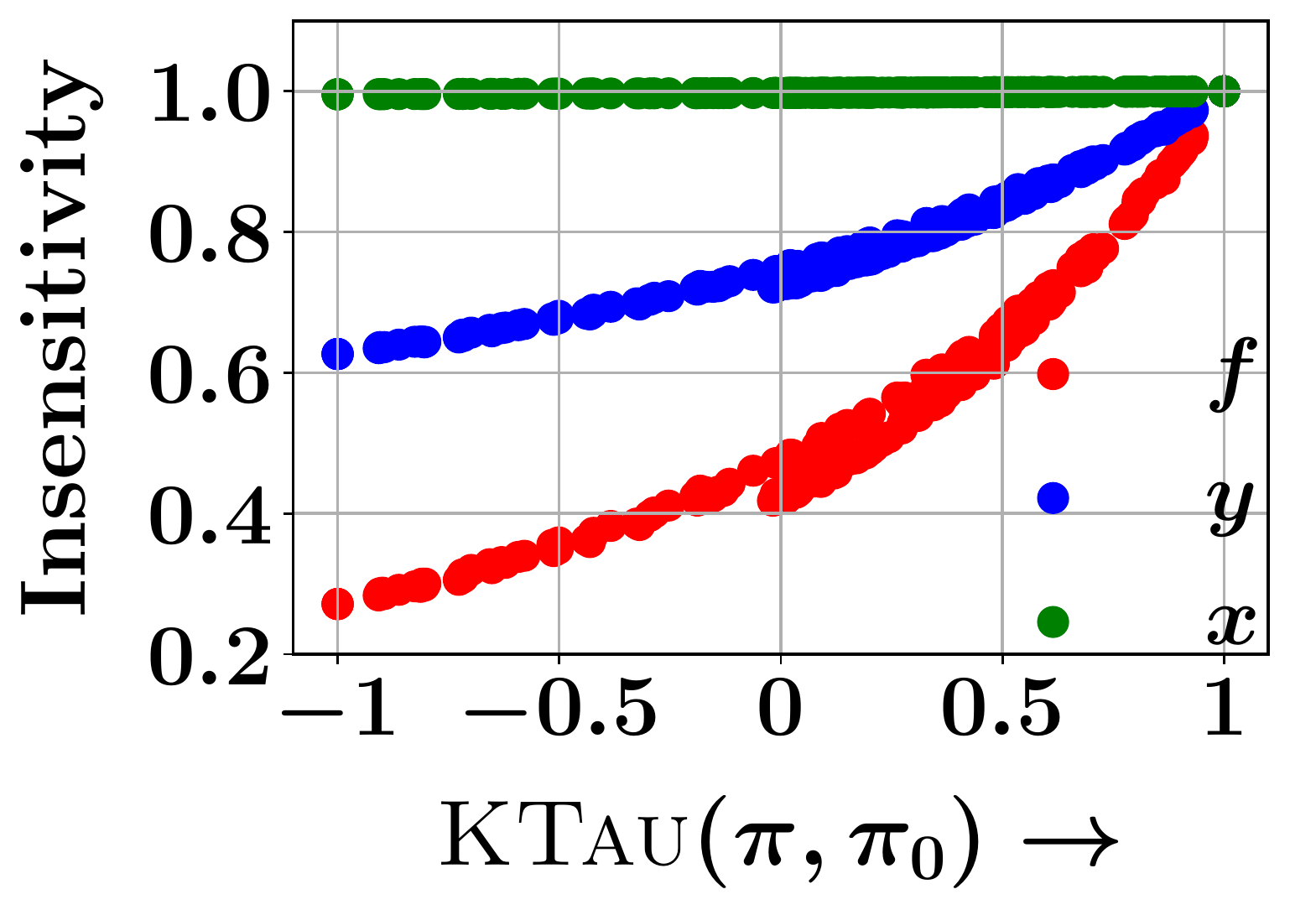}}\hspace{2mm}
\subfloat[Citeseer]{\includegraphics[width=.20\textwidth]{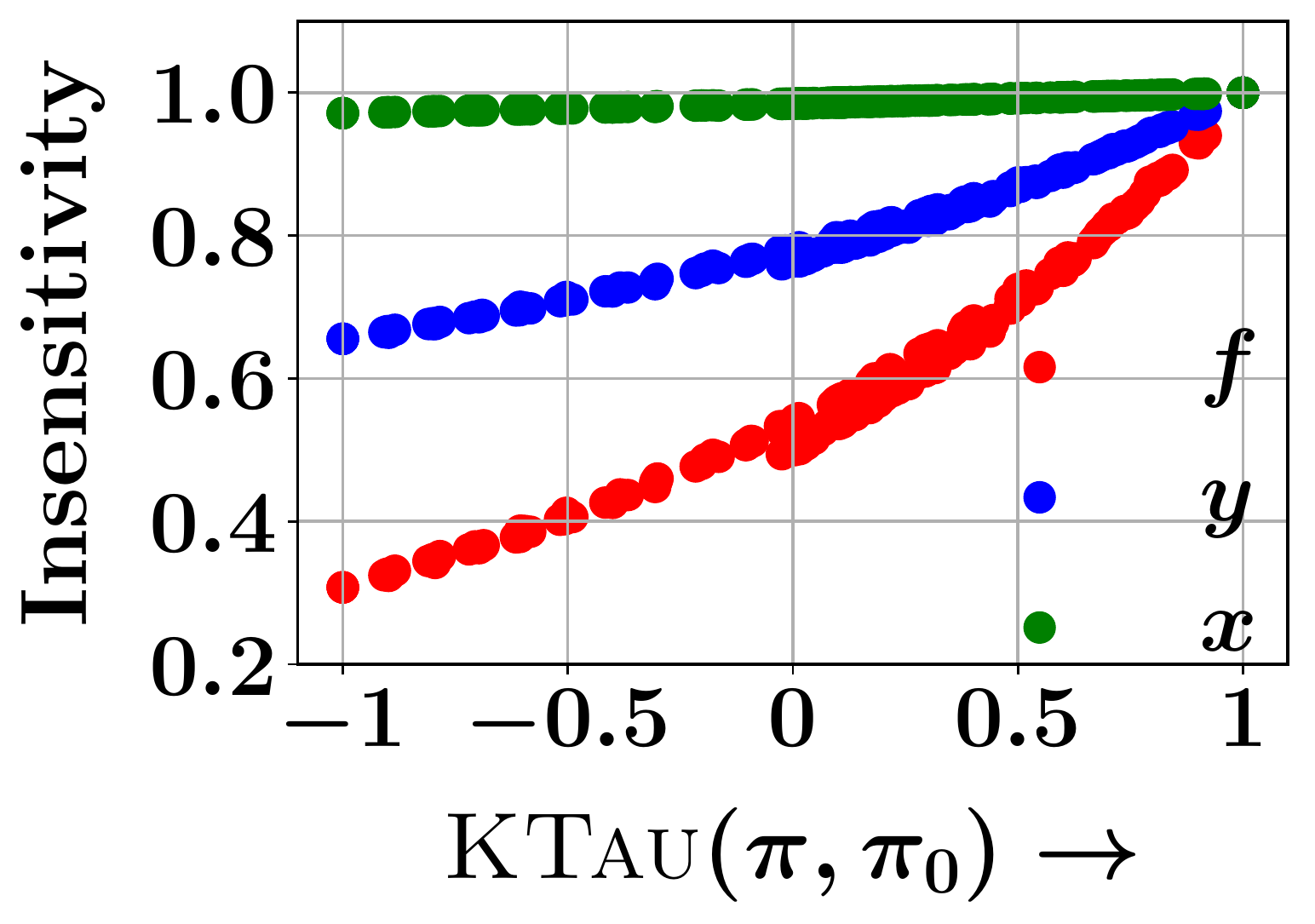}}

\caption{Insensitivity of neighborhood features $\set{\fb_v \,|\, v\in\nbr(u)}$ LSTM output $\{\yb\}$ and the 
resultant node embeddings $\xb_u$ with respect to neighbor order permutations.}
\label{fig:KtauVsInsensitivity}
\end{figure}

To probe this phenomenon, we instrument the stability of $\fb, \yb, \xb$ to different permutations. Specifically, we define $\text{insensitivity}(\zb;\pib,\pib_0)=\sum_{u\in V}\textsc{sim}(\zb_{u;\pib},\zb_{u;\pib_0})/|V|$ for any vector or sequence $\zb$.  We compute insensitivity of the input sequence $\set{\fb_v : v\in\nbr(u)}$, the intermediate LSTM output $\set{\yb}$ and the final embedding $\xb_u$ with respect to different permutations $\pib$. Figure~\ref{fig:KtauVsInsensitivity} summarizes the results, and shows that as information flows through \our{} stages, from input feature sequence to the final embeddings, the insensitivity of the underlying signals increases.  Thus, our adversarial training smoothly turns permutation-sensitive input sequences into permutation invariant node embeddings, without any explicit symmetric aggregator.

\begin{figure}[t]
\centering
\subfloat[Tensorized]{ \includegraphics[width=0.40\hsize]{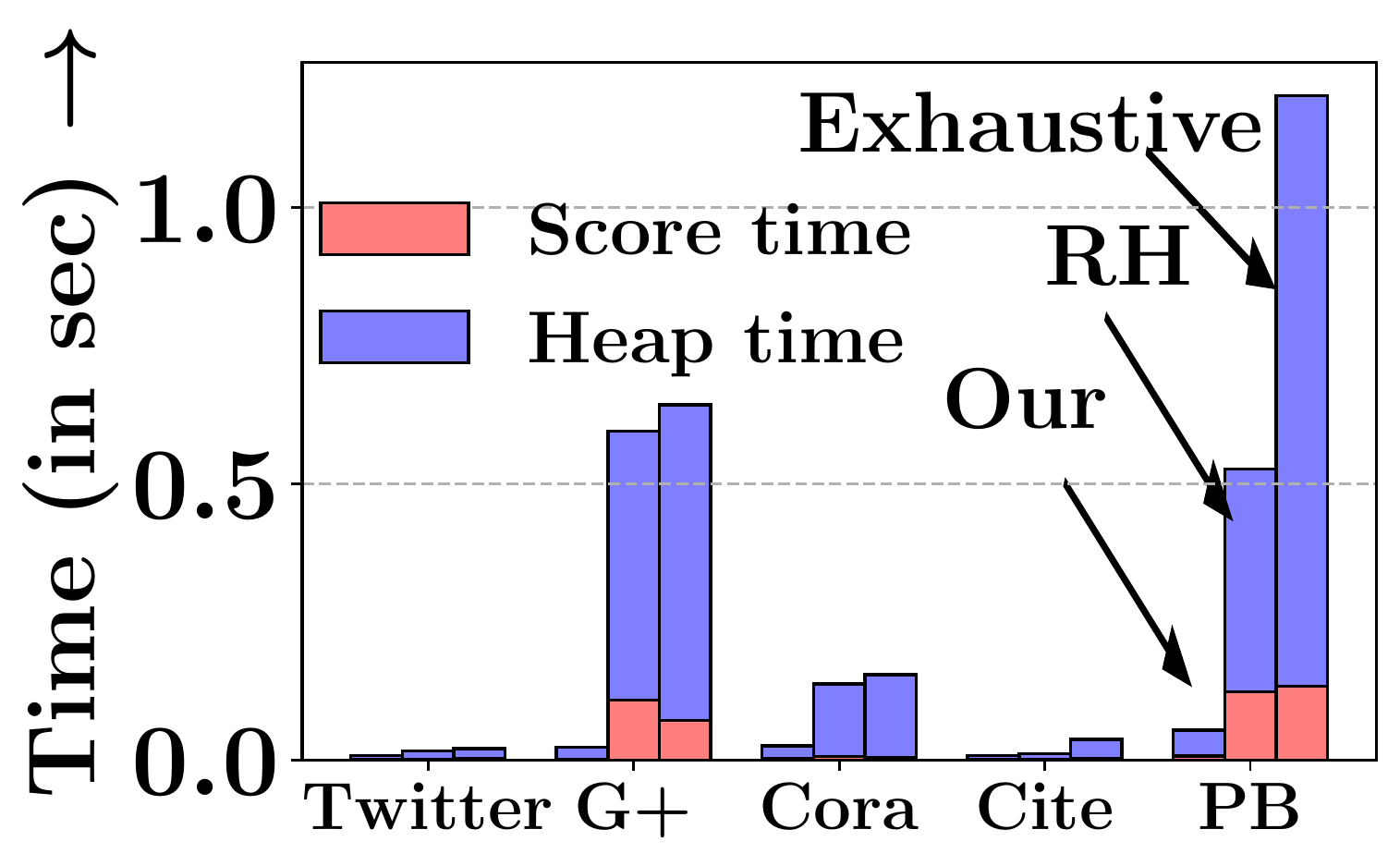}}\hspace{2mm}
\subfloat[Non-tensorized]{ \includegraphics[width=0.40\hsize]{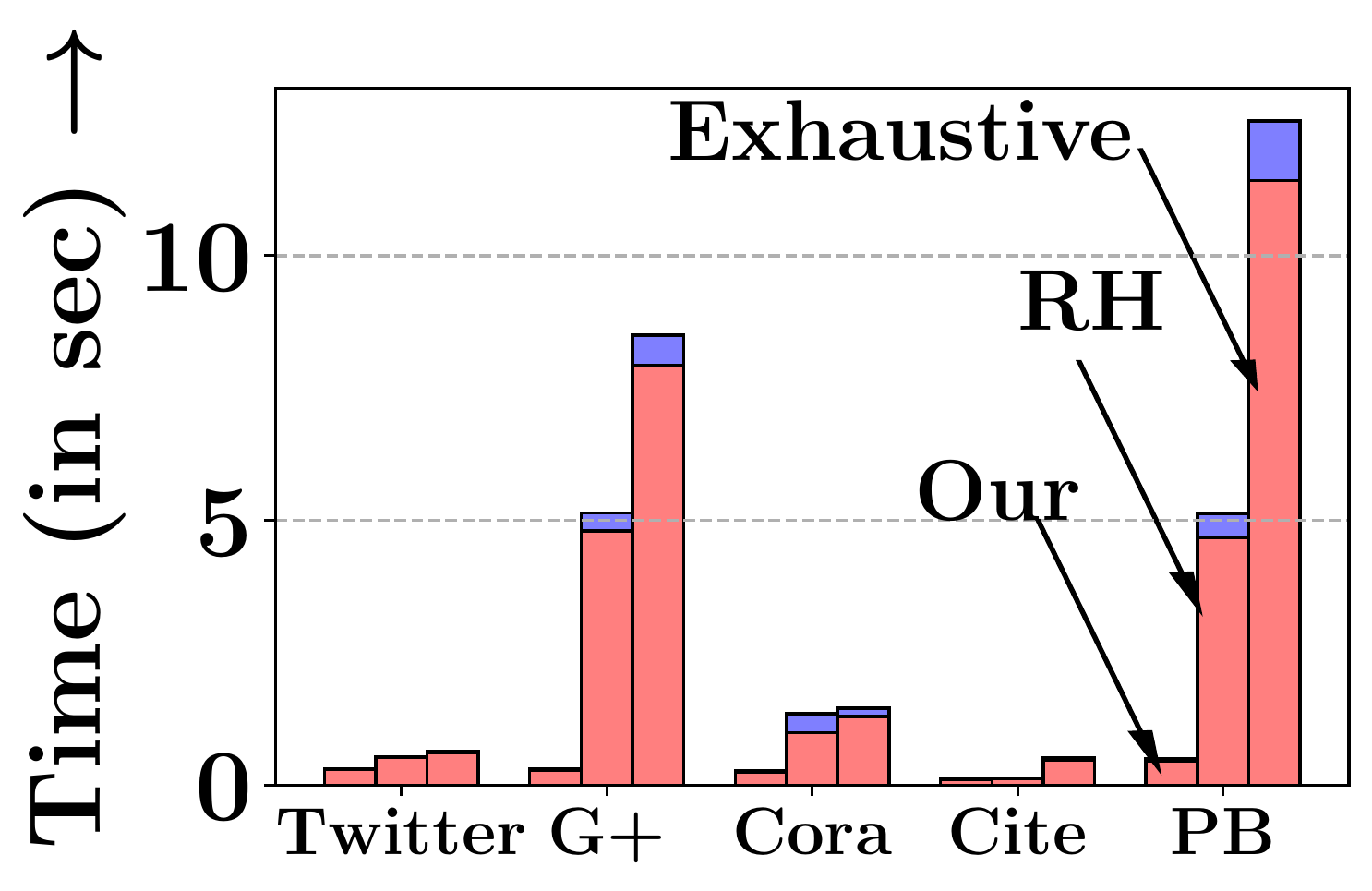}}  

\caption{{Running time for our LSH based scalable prediction, random-hyperplane based LSH method, exhaustive comparison.
}}
\label{fig:TopkTimeHashing}
\end{figure}

\subsection{Performance of Hashing Methods}

Finally, we address \textbf{RQ4} by studying the performance of our LSH method (Section~\ref{sec:HashOpt}).  Specifically, we compare the time spent in similarity computation and heap operations of our hashing method against random hyperplane based hashing (Section~\ref{sec:Hyperplanes}), compared to exhaustive computation of pairwise scores (as a slow but ``relatively perfect'' baseline).  Since vectorized similarity computation inside Torch may be faster than numpy, we provide results on both implementations.  
Figure~\ref{fig:TopkTimeHashing} summarizes results in terms of running time.  It shows that: (1)~hashing using $C_\psi$ leads to considerable savings in reporting top-$K$ node-pairs with respect to both random hyperplane based hashing {and exhaustive enumeration}, and (2)~the gains increase with increasing graph sizes (from \google{} to PB).
Because LSH-based top-$K$ retrieval may discard relevant nodes after $K$, it is more appropriate to study ranking degradation in terms of decrease in NDCG (rather than MAP).  Suppose we insist that NDCG be at least 85, 90, or 95\% of exhaustive NDCG.  How selective is a hashing strategy, in terms of the factor of query speedup (because of buckets pruned in Algorithm~\ref{algo:LshTopK})?  Table~\ref{tab:LshNdcg} shows that our hashing method provides better pruning than random hyperplane for a given level of NDCG degradation.

\begin{table}[ht]
\centering

\centering
\maxsizebox{\hsize}{!}{ \tabcolsep 2pt 
\begin{tabular}{|c|c|c|c|c|c|c|}
\hline
    & \multicolumn{6}{|c|}{Minimum NDCG as \% of exhaustive NDCG} 
    \\ \cline{2-7} 
    & \multicolumn{2}{|c|}{85\%} & \multicolumn{2}{c|}{90\%} & \multicolumn{2}{c|}{95\%}    \\ \cline{2-7} 
    & Twitter     & \google    & Twitter     & \google    & \multicolumn{1}{c|}{Twitter} & \google \\ \hline
Our Hashing  & 6.67        & 12.5      & 6.67        & 10        & \multicolumn{1}{c|}{6.25}    & 5.5    \\ \hline
RH  & 1.78        & 3.45      & 1.78        & 3.45      & \multicolumn{1}{c|}{1.78}    & 3.45  \\ \hline
\end{tabular} }

\caption{Speedup achieved by different hashing methods under various permitted NDCG degradation limits. }
\label{tab:LshNdcg}
\end{table}

\section{Conclusion}
\label{sec:End}
We presented \our, a novel LP formulation that combines a recurrent, order-sensitive graph neighbor aggregator with an adversarial generator of neighbor permutations.  \our{} achieves LP accuracy comparable to or better than sampling a number of permutations by brute force, and is faster to train.  \our{} is also superior to a number of LP baselines.  In addition, we formulate an optimization to map \our's node embeddings to a suitable locality-sensitive hash, which greatly speeds up reporting of the most likely edges. It would be interesting to extend \our{} to other downstream network analyses, \eg, node classification, community detection, or knowledge graph completion.

\section*{Acknowledgements}
Partly supported by an IBM AI Horizons Grant. Thanks to Chitrank Gupta and Yash Jain for helping rectify an error in an earlier evaluation method.


\begingroup
\bibliography{refs,voila}

\begin{thebibliography}{53}
\providecommand{\natexlab}[1]{#1}
\providecommand{\url}[1]{\texttt{#1}}
\providecommand{\urlprefix}{URL }
\expandafter\ifx\csname urlstyle\endcsname\relax
  \providecommand{\doi}[1]{doi:\discretionary{}{}{}#1}\else
  \providecommand{\doi}{doi:\discretionary{}{}{}\begingroup
  \urlstyle{rm}\Url}\fi

\bibitem[{Ackland et~al.(2005)}]{ackland2005mapping}
Ackland, R.; et~al. 2005.
\newblock Mapping the US political blogosphere: Are conservative bloggers more
  prominent?
\newblock In \emph{BlogTalk Downunder 2005 Conference, Sydney}. BlogTalk
  Downunder 2005 Conference, Sydney.

\bibitem[{Adamic and Adar(2003)}]{AdamicA2003FriendsNeighbors}
Adamic, L.~A.; and Adar, E. 2003.
\newblock Friends and neighbors on the {Web}.
\newblock \emph{Social Networks} 25(3): 211 -- 230.
\newblock ISSN 0378-8733.
\newblock \doi{http://dx.doi.org/10.1016/S0378-8733(03)00009-1}.
\newblock
  \urlprefix\url{http://pkudlib.org/qmeiCourse/files/FriendsAndNeighbors.pdf}.

\bibitem[{Backstrom and Leskovec(2011)}]{BackstromL2011SRW}
Backstrom, L.; and Leskovec, J. 2011.
\newblock Supervised random walks: predicting and recommending links in social
  networks.
\newblock In \emph{WSDM Conference}, 635--644.
\newblock
  \urlprefix\url{http://cs.stanford.edu/people/jure/pubs/linkpred-wsdm11.pdf}.

\bibitem[{Bloem-Reddy and Teh(2019)}]{bloem2019probabilistic}
Bloem-Reddy, B.; and Teh, Y.~W. 2019.
\newblock Probabilistic symmetry and invariant neural networks.
\newblock \emph{arXiv preprint arXiv:1901.06082} .

\bibitem[{Borovkova and Tsiamas(2019)}]{borovkova2019ensemble}
Borovkova, S.; and Tsiamas, I. 2019.
\newblock An ensemble of LSTM neural networks for high-frequency stock market
  classification.
\newblock \emph{Journal of Forecasting} 38(6): 600--619.

\bibitem[{Charikar(2002)}]{Charikar2002lsh}
Charikar, M.~S. 2002.
\newblock Similarity estimation techniques from rounding algorithms.
\newblock In \emph{STOC}, 380--388.
\newblock \urlprefix\url{https://dl.acm.org/doi/pdf/10.1145/509907.509965}.

\bibitem[{Cohen-Karlik, David, and
  Globerson(2020)}]{cohenkarlik2020regularizing}
Cohen-Karlik, E.; David, A.~B.; and Globerson, A. 2020.
\newblock Regularizing Towards Permutation Invariance in Recurrent Models.
\newblock In \emph{NeurIPS}.
\newblock \urlprefix\url{https://arxiv.org/abs/2010.13055}.

\bibitem[{Cuturi(2013)}]{Cuturi2013sinkhorn}
Cuturi, M. 2013.
\newblock Sinkhorn distances: Lightspeed computation of optimal transport.
\newblock In \emph{NeurIPS}, 2292--2300.
\newblock
  \urlprefix\url{https://papers.nips.cc/paper/4927-sinkhorn-distances-lightspeed-computation-of-optimal-transport.pdf}.

\bibitem[{Garg, Jegelka, and Jaakkola(2020)}]{garg2020generalization}
Garg, V.~K.; Jegelka, S.; and Jaakkola, T. 2020.
\newblock Generalization and representational limits of graph neural networks.
\newblock \emph{arXiv preprint arXiv:2002.06157} .

\bibitem[{Getoor(2005)}]{getoor2005link}
Getoor, L. 2005.
\newblock Link-based classification.
\newblock In \emph{Advanced methods for knowledge discovery from complex data},
  189--207. Springer.

\bibitem[{Gionis, Indyk, and Motwani(1999)}]{GionisIM1999hash}
Gionis, A.; Indyk, P.; and Motwani, R. 1999.
\newblock Similarity Search in High Dimensions via Hashing.
\newblock In \emph{VLDB Conference}, 518--529.
\newblock See \url{http://citeseer.nj.nec.com/gionis97similarity.html}.

\bibitem[{Grover and Leskovec(2016)}]{grover2016node2vec}
Grover, A.; and Leskovec, J. 2016.
\newblock node2vec: Scalable feature learning for networks.
\newblock In \emph{SIGKDD}.

\bibitem[{Hamilton, Ying, and Leskovec(2017)}]{hamilton2017inductive}
Hamilton, W.; Ying, Z.; and Leskovec, J. 2017.
\newblock Inductive representation learning on large graphs.
\newblock In \emph{Advances in neural information processing systems},
  1024--1034.

\bibitem[{Hochreiter and Schmidhuber(1997)}]{HochreiterS1997LSTM}
Hochreiter, S.; and Schmidhuber, J. 1997.
\newblock Long Short-Term Memory.
\newblock \emph{Neural Computation} 9(8): 1735--1780.
\newblock
  \urlprefix\url{https://www.mitpressjournals.org/doi/pdfplus/10.1162/neco.1997.9.8.1735}.

\bibitem[{Joachims(2005)}]{Joachims2005multivariate}
Joachims, T. 2005.
\newblock A support vector method for multivariate performance measures.
\newblock In \emph{ICML}, 377--384.
\newblock ISBN 1-59593-180-5.
\newblock \doi{http://doi.acm.org/10.1145/1102351.1102399}.
\newblock
  \urlprefix\url{http://www.machinelearning.org/proceedings/icml2005/papers/048_ASupport_Joachims.pdf}.

\bibitem[{Katz(1997)}]{Katz1997start}
Katz, B. 1997.
\newblock From Sentence Processing to Information Access on the {World} Wide
  {Web}.
\newblock In \emph{AAAI Spring Symposium on Natural Language Processing for the
  World Wide Web}, 77--94. Stanford CA: Stanford University.
\newblock See \url{http://www.ai.mit.edu/people/boris/webaccess/}.

\bibitem[{Kipf and Welling(2016{\natexlab{a}})}]{kipf2016semi}
Kipf, T.~N.; and Welling, M. 2016{\natexlab{a}}.
\newblock Semi-supervised classification with graph convolutional networks.
\newblock \emph{arXiv preprint arXiv:1609.02907} .

\bibitem[{Kipf and Welling(2016{\natexlab{b}})}]{kipf2016variational}
Kipf, T.~N.; and Welling, M. 2016{\natexlab{b}}.
\newblock Variational graph auto-encoders.
\newblock \emph{arXiv preprint arXiv:1611.07308} .

\bibitem[{Kulis and Darrell(2009)}]{KulisD2009HashLearn}
Kulis, B.; and Darrell, T. 2009.
\newblock Learning to hash with binary reconstructive embeddings.
\newblock In \emph{NeurIPS}, 1042--1050.
\newblock
  \urlprefix\url{http://papers.nips.cc/paper/3667-learning-to-hash-with-binary-reconstructive-embeddings.pdf}.

\bibitem[{Lee et~al.(2019)Lee, Lee, Kim, Kosiorek, Choi, and Teh}]{lee2019set}
Lee, J.; Lee, Y.; Kim, J.; Kosiorek, A.; Choi, S.; and Teh, Y.~W. 2019.
\newblock Set transformer: A framework for attention-based
  permutation-invariant neural networks.
\newblock In \emph{ICML}.

\bibitem[{Leskovec et~al.(2010)Leskovec, Chakrabarti, Kleinberg, Faloutsos, and
  Ghahramani}]{leskovec2010kronecker}
Leskovec, J.; Chakrabarti, D.; Kleinberg, J.; Faloutsos, C.; and Ghahramani, Z.
  2010.
\newblock Kronecker graphs: An approach to modeling networks.
\newblock \emph{Journal of Machine Learning Research} 11(Feb): 985--1042.

\bibitem[{Leskovec and Mcauley(2012)}]{leskovec2012learning}
Leskovec, J.; and Mcauley, J.~J. 2012.
\newblock Learning to discover social circles in ego networks.
\newblock In \emph{NeuIPS}.

\bibitem[{Liben-Nowell and Kleinberg(2007)}]{LibenNowellK2007LinkPred}
Liben-Nowell, D.; and Kleinberg, J. 2007.
\newblock The link-prediction problem for social networks.
\newblock \emph{Journal of the American Society for Information Science and
  Technology} 58(7): 1019--1031.
\newblock ISSN 1532-2890.
\newblock \doi{10.1002/asi.20591}.
\newblock
  \urlprefix\url{https://onlinelibrary.wiley.com/doi/full/10.1002/asi.20591}.

\bibitem[{Lichtenwalter, Lussier, and
  Chawla(2010)}]{LichtenwalterLC2010PropFlow}
Lichtenwalter, R.~N.; Lussier, J.~T.; and Chawla, N.~V. 2010.
\newblock New perspectives and methods in link prediction.
\newblock In \emph{SIGKDD Conference}, 243--252. Washington, DC, USA: ACM.
\newblock ISBN 978-1-4503-0055-1.
\newblock \doi{10.1145/1835804.1835837}.
\newblock
  \urlprefix\url{http://users.cs.fiu.edu/~lzhen001/activities/KDD_USB_key_2010/docs/p243.pdf}.

\bibitem[{Liu et~al.(2012)Liu, Wang, Ji, Jiang, and Chang}]{Liu+2012SupHash}
Liu, W.; Wang, J.; Ji, R.; Jiang, Y.-G.; and Chang, S.-F. 2012.
\newblock Supervised hashing with kernels.
\newblock In \emph{IEEE CVPR}, 2074--2081.
\newblock
  \urlprefix\url{https://ieeexplore.ieee.org/stamp/stamp.jsp?arnumber=6247912}.

\bibitem[{Mena et~al.(2018)Mena, Belanger, Linderman, and
  Snoek}]{Mena+2018GumbelSinkhorn}
Mena, G.; Belanger, D.; Linderman, S.; and Snoek, J. 2018.
\newblock Learning latent permutations with gumbel-sinkhorn networks.
\newblock \emph{arXiv preprint arXiv:1802.08665}
  \urlprefix\url{https://arxiv.org/pdf/1802.08665.pdf}.

\bibitem[{Murphy et~al.(2019{\natexlab{a}})Murphy, Srinivasan, Rao, and
  Ribeiro}]{murphy2019janossy}
Murphy, R.~L.; Srinivasan, B.; Rao, V.; and Ribeiro, B. 2019{\natexlab{a}}.
\newblock Janossy pooling: Learning deep permutation-invariant functions for
  variable-size inputs.
\newblock \emph{ICLR} \urlprefix\url{https://arxiv.org/pdf/1811.01900}.

\bibitem[{Murphy et~al.(2019{\natexlab{b}})Murphy, Srinivasan, Rao, and
  Ribeiro}]{murphy2019relational}
Murphy, R.~L.; Srinivasan, B.; Rao, V.; and Ribeiro, B. 2019{\natexlab{b}}.
\newblock Relational pooling for graph representations.
\newblock \emph{arXiv preprint arXiv:1903.02541} .

\bibitem[{NT and Maehara(2019)}]{nt2019revisiting}
NT, H.; and Maehara, T. 2019.
\newblock Revisiting graph neural networks: All we have is low-pass filters.
\newblock \emph{arXiv preprint arXiv:1905.09550} .

\bibitem[{Pabbaraju and Jain(2019)}]{PabbarajuJ2019permute}
Pabbaraju, C.; and Jain, P. 2019.
\newblock Learning Functions over Sets via Permutation Adversarial Networks.
\newblock \emph{arXiv preprint arXiv:1907.05638}
  \urlprefix\url{https://arxiv.org/pdf/1907.05638}.

\bibitem[{Perozzi, Al-Rfou, and Skiena(2014)}]{perozzi2014deepwalk}
Perozzi, B.; Al-Rfou, R.; and Skiena, S. 2014.
\newblock Deepwalk: Online learning of social representations.
\newblock In \emph{KDD}, 701--710.

\bibitem[{Qi et~al.(2017)Qi, Su, Mo, and Guibas}]{qi2017pointnet}
Qi, C.~R.; Su, H.; Mo, K.; and Guibas, L.~J. 2017.
\newblock Pointnet: Deep learning on point sets for 3d classification and
  segmentation.
\newblock In \emph{Proceedings of the IEEE conference on computer vision and
  pattern recognition}, 652--660.

\bibitem[{Ravanbakhsh, Schneider, and Poczos(2016)}]{ravanbakhsh2016deep}
Ravanbakhsh, S.; Schneider, J.; and Poczos, B. 2016.
\newblock Deep learning with sets and point clouds.
\newblock \emph{arXiv preprint arXiv:1611.04500} .

\bibitem[{Salha et~al.(2019)Salha, Limnios, Hennequin, Tran, and
  Vazirgiannis}]{Salha+2019gravity}
Salha, G.; Limnios, S.; Hennequin, R.; Tran, V.-A.; and Vazirgiannis, M. 2019.
\newblock Gravity-Inspired Graph Autoencoders for Directed Link Prediction.
\newblock In \emph{CIKM}, 589–598.
\newblock \urlprefix\url{https://doi.org/10.1145/3357384.3358023}.

\bibitem[{Sarkar, Chakrabarti, and Moore(2011)}]{SarkarCM2011LPembed}
Sarkar, P.; Chakrabarti, D.; and Moore, A.~W. 2011.
\newblock Theoretical justification of popular link prediction heuristics.
\newblock In \emph{COLT}.

\bibitem[{Schlichtkrull et~al.(2018)Schlichtkrull, Kipf, Bloem, Van Den~Berg,
  Titov, and Welling}]{schlichtkrull2018rgcn}
Schlichtkrull, M.; Kipf, T.~N.; Bloem, P.; Van Den~Berg, R.; Titov, I.; and
  Welling, M. 2018.
\newblock Modeling relational data with graph convolutional networks.
\newblock In \emph{European Semantic Web Conference}, 593--607.
\newblock \urlprefix\url{https://arxiv.org/pdf/1703.06103}.

\bibitem[{Sen et~al.(2008)Sen, Namata, Bilgic, Getoor, Galligher, and
  Eliassi-Rad}]{sen2008collective}
Sen, P.; Namata, G.; Bilgic, M.; Getoor, L.; Galligher, B.; and Eliassi-Rad, T.
  2008.
\newblock Collective classification in network data.
\newblock \emph{AI magazine} 29(3): 93--93.

\bibitem[{Shi, Oliva, and Niethammer(2020)}]{shi2020deep}
Shi, Y.; Oliva, J.; and Niethammer, M. 2020.
\newblock Deep Message Passing on Sets.
\newblock In \emph{AAAI}, 5750--5757.

\bibitem[{Sinkhorn(1967)}]{Sinkhorn1967diagonal}
Sinkhorn, R. 1967.
\newblock Diagonal equivalence to matrices with prescribed row and column sums.
\newblock \emph{The American Mathematical Monthly} 74(4): 402--405.
\newblock \urlprefix\url{https://www.jstor.org/stable/pdf/2314570.pdf}.

\bibitem[{Skianis et~al.(2020)Skianis, Nikolentzos, Limnios, and
  Vazirgiannis}]{skianis2020rep}
Skianis, K.; Nikolentzos, G.; Limnios, S.; and Vazirgiannis, M. 2020.
\newblock Rep the set: Neural networks for learning set representations.
\newblock In \emph{International Conference on Artificial Intelligence and
  Statistics}, 1410--1420. PMLR.

\bibitem[{Stelzner, Kersting, and Kosiorek(2020)}]{stelznergenerative}
Stelzner, K.; Kersting, K.; and Kosiorek, A.~R. 2020.
\newblock Generative Adversarial Set Transformers.
\newblock In \emph{Workshop on Object-Oriented Learning at ICML~2020}.
\newblock
  \urlprefix\url{https://www.ml.informatik.tu-darmstadt.de/papers/stelzner2020ood_gast.pdf}.

\bibitem[{Tang et~al.(2015)Tang, Qu, Wang, Zhang, Yan, and Mei}]{tang2015line}
Tang, J.; Qu, M.; Wang, M.; Zhang, M.; Yan, J.; and Mei, Q. 2015.
\newblock {LINE}: Large-scale information network embedding.
\newblock In \emph{WWW Conference}, 1067--1077.

\bibitem[{Veli{\v{c}}kovi{\'c} et~al.(2017)Veli{\v{c}}kovi{\'c}, Cucurull,
  Casanova, Romero, Lio, and Bengio}]{velivckovic2017graph}
Veli{\v{c}}kovi{\'c}, P.; Cucurull, G.; Casanova, A.; Romero, A.; Lio, P.; and
  Bengio, Y. 2017.
\newblock Graph attention networks.
\newblock \emph{arXiv preprint arXiv:1710.10903} .

\bibitem[{Wagstaff et~al.(2019)Wagstaff, Fuchs, Engelcke, Posner, and
  Osborne}]{wagstaff2019limitations}
Wagstaff, E.; Fuchs, F.~B.; Engelcke, M.; Posner, I.; and Osborne, M. 2019.
\newblock On the limitations of representing functions on sets.
\newblock \emph{arXiv preprint arXiv:1901.09006} .

\bibitem[{Wang et~al.(2019)Wang, Ren, He, Zhang, and Hu}]{wang2019robust}
Wang, Z.; Ren, Z.; He, C.; Zhang, P.; and Hu, Y. 2019.
\newblock Robust Embedding with Multi-Level Structures for Link Prediction.
\newblock In \emph{IJCAI}, 5240--5246.
\newblock \urlprefix\url{https://www.ijcai.org/Proceedings/2019/0728.pdf}.

\bibitem[{Weiss, Torralba, and Fergus(2009)}]{WeissTF2009SpectralHashing}
Weiss, Y.; Torralba, A.; and Fergus, R. 2009.
\newblock Spectral hashing.
\newblock In \emph{NeurIPS}, 1753--1760.
\newblock
  \urlprefix\url{https://papers.nips.cc/paper/3383-spectral-hashing.pdf}.

\bibitem[{Wu et~al.(2019)Wu, Zhang, Souza~Jr, Fifty, Yu, and
  Weinberger}]{wu2019simplifying}
Wu, F.; Zhang, T.; Souza~Jr, A. H.~d.; Fifty, C.; Yu, T.; and Weinberger, K.~Q.
  2019.
\newblock Simplifying graph convolutional networks.
\newblock \emph{arXiv preprint arXiv:1902.07153} .

\bibitem[{Xu et~al.(2018{\natexlab{a}})Xu, Hu, Leskovec, and
  Jegelka}]{xu2018powerful}
Xu, K.; Hu, W.; Leskovec, J.; and Jegelka, S. 2018{\natexlab{a}}.
\newblock How powerful are graph neural networks?
\newblock \emph{arXiv preprint arXiv:1810.00826} .

\bibitem[{Xu et~al.(2018{\natexlab{b}})Xu, Li, Tian, Sonobe, Kawarabayashi, and
  Jegelka}]{xu2018representation}
Xu, K.; Li, C.; Tian, Y.; Sonobe, T.; Kawarabayashi, K.-i.; and Jegelka, S.
  2018{\natexlab{b}}.
\newblock Representation learning on graphs with jumping knowledge networks.
\newblock \emph{arXiv preprint arXiv:1806.03536} .

\bibitem[{Yadati et~al.(2018)Yadati, Nitin, Nimishakavi, Yadav, Louis, and
  Talukdar}]{yadati2018hyperlp}
Yadati, N.; Nitin, V.; Nimishakavi, M.; Yadav, P.; Louis, A.; and Talukdar, P.
  2018.
\newblock Link prediction in hypergraphs using graph convolutional networks.
\newblock Manuscript.
\newblock \urlprefix\url{https://openreview.net/forum?id=ryeaZhRqFm}.

\bibitem[{You, Ying, and Leskovec(2019)}]{you2019position}
You, J.; Ying, R.; and Leskovec, J. 2019.
\newblock Position-aware graph neural networks.
\newblock \emph{arXiv preprint arXiv:1906.04817} .

\bibitem[{Zaheer et~al.(2017)Zaheer, Kottur, Ravanbakhsh, Poczos,
  Salakhutdinov, and Smola}]{zaheer2017deep}
Zaheer, M.; Kottur, S.; Ravanbakhsh, S.; Poczos, B.; Salakhutdinov, R.~R.; and
  Smola, A.~J. 2017.
\newblock Deep sets.
\newblock In \emph{Advances in neural information processing systems},
  3391--3401.

\bibitem[{Zhang and Chen(2018)}]{ZhangC2018LinkPredGNN}
Zhang, M.; and Chen, Y. 2018.
\newblock Link prediction based on graph neural networks.
\newblock In \emph{NeurIPS}.

\end{thebibliography}
\endgroup

\newpage
\appendix
\onecolumn
\begingroup \centering \bfseries \LARGE \ztitle \\
(Appendix) \par\smallskip
\endgroup

\section*{Contents}
\begin{itemize}
\item In Appendix~\ref{sec:Rel} we provide a more detailed discussion of prior work.
\item In Appendix~\ref{sec:Hashing} we present additional details about our hashing and bucketing methods.
\item In Appendix~\ref{sec:ExptSetupDetail} we give the specifications of all the network modules used in \our and complete settings of our experiments, which, together with our code, makes our results reproducible.
\end{itemize}

\section{Detailed commentary on prior work}
\label{sec:Rel}

\paragraph{\bfseries Link prediction and GNNs}
Unsupervised LP algorithms compute a heuristic confidence score of a potential edge, given a node pair, based solely on local network structures \citep{AdamicA2003FriendsNeighbors, LibenNowellK2007LinkPred}.  Adamic-Adar (AA), common-neighbor (CN) and Jaccard coefficient (JC) are examples.  Prior to deep learning, conventional supervised learning was successfully used for LP \citep{LichtenwalterLC2010PropFlow, BackstromL2011SRW, Katz1997start}.

Recent years have witnessed a surge of interest in modeling and learning latent node features, called node embeddings or representations.  These are low dimensional compact vectors, compressed from the high dimensional neighborhood information of the larger graph. In contrast to hand-engineered features, node embeddings are modeled using highly expressive neural networks that are trained using the observable graph structure.

Node2Vec~\cite{grover2016node2vec}, DeepWalk~\cite{perozzi2014deepwalk} and LINE~\cite{tang2015line} were among the earliest attempts to fit node embeddings. GCNs \citep{kipf2016semi} and RGCNs \citep{schlichtkrull2018rgcn} soon followed. \citet{wang2019robust} exploited multi-level graph coarsening in their proposed system called MGNN, which benefits from naturally hierarhical knowledge graphs (KGs). \citet{Salha+2019gravity} extended the GCN paradigm to directed graphs. \citet{yadati2018hyperlp} extended GCNs to hypergraphs.  Other notable enhancements were proposed as GraphSAGE~\cite{hamilton2017inductive}, GAT~\cite{velivckovic2017graph}, SEAL~\cite{ZhangC2018LinkPredGNN}, GIN~\cite{xu2018powerful}, JKN~\cite{xu2018representation}, and P-GNN~\cite{you2019position}, \emph{inter alia}.

\paragraph{\bfseries Neural permutation gadgets}
\citet{Sinkhorn1967diagonal} used iterative row and column scaling as an effective way to impute matrices, given marginal constraints.  \citet{Cuturi2013sinkhorn} exploited this to solve transportation problems approximately.  It was soon realized \citep{Mena+2018GumbelSinkhorn, PabbarajuJ2019permute} that row and column scaling transform an arbitrary matrix to a near-permutation matrix, while allowing backpropagation.  After the seminal deep sets work of \citet{zaheer2017deep}, several efforts \citep{lee2019set, bloem2019probabilistic, shi2020deep, stelznergenerative} were made to capture dependencies between set elements while retaining order invariance by design \citep{murphy2019janossy}.

\paragraph{\bfseries (Supervised) locality-sensitive hashing}
LSH was proposed in path-breaking papers by \citet{GionisIM1999hash} and \citet{Charikar2002lsh}.  These were \emph{data-oblivious} hashing protocols.  Later, data-driven, supervised hashing approaches
\cite{KulisD2009HashLearn, WeissTF2009SpectralHashing, Liu+2012SupHash} were proposed.


\section{Additional details about proposed hashing method}
\label{sec:Hashing}

We form the buckets using the recipe of \citet{GionisIM1999hash}, summarized here for completeness.  Given hash bit positions $1,\ldots,H$, we select $J<H$ bit positions uniformly at random, $L$ times.  ($J,L$ are chosen based on $N$ and performance targets.)  Let these bit indices be $I_1,\ldots,I_\ell,\ldots,I_L$ and let $g_{u,\ell} = \bm{b}_u\!\downarrow_{I_\ell}$ be the hash code of node $u$, projected to the bit positions~$I_\ell$.  We thus obtain $L$ bitvectors from $\xb_u$, called $g_{u,1},\ldots,g_{u,L} \in \set{-1,+1}^J$, which represents a number in $[0,2^J-1]$.  There are $L$ hashtables, each with $2^J$ buckets.
Node $u$ is registered in each hashtable once. In hashtable number $\ell$, it goes into the bucket numbered~$g_{u,\ell}$.  Qualitatively, if nodes $u$ and $v$ occupy the same bucket in many of the $L$ hashtables, they are very similar.  We score $\set{u,v}$ if they share a bucket in any of the $L$ hashtables.

In practice, we set $J{=}8$ and $L{=}10$.
The hashing/compression network $C_{\psi}$ is devised with a single linear layer of dimension $(D,H)$. We choose the output hashcode dimension ($H$) to be same as input embedding dimension ($D$), i.e.,~16.

%


%

\section{Additional details on experimental setup}
\label{sec:ExptSetupDetail}

\subsection{Design specifications of \our}
\label{sec:NetDetails}
Excluding the hashing machinery, \our has three neural modules: 
\begin{enumerate*}
    \item The LSTM aggregator $\text{LSTM}_{\theta}$ in Eq.~\eqref{eq:lstm}.
    \item The nonlinear component in the outer layer $\sigma_{\theta}$ in Eq.~\eqref{eq:sigma-theta-intro}.
    \item The permutation generator network $\Tb_{\phi}$ in Eq.~\eqref{eq:TIntro}.
\end{enumerate*}
In the following, we describe the specifications of these components, beginning with the node features $\set{\fb_u}$.

\paragraph{Specification of $\fb_\bullet$}
For Cora and Citeseer datasets, node features $\set{\fb_u}$ are binary vectors indicating presence/absence of corresponding keywords in the document. For the remaining datasets, we define node features as the one-hot representations of the unique node labels.

\paragraph{Specification of $\text{LSTM}_{\theta}$} Across all experiments we used an LSTM with hidden size $32$.

\paragraph{Specification of $\sigma_{\theta}$} 
We design $\sigma_{\theta}$ (Eq.~\eqref{eq:sigma-theta-intro}) with a fully connected single layer feed forward network on top of the LSTM. This outputs the final node embeddings with $\text{dim}(\xb_\bullet) = 16$.

\paragraph{Specification of $\Tb_{\phi}$ } 
We design $\Tb_{\phi}$  (Eq.~\eqref{eq:TIntro}) using a three layer neural network which consists of one linear, one ReLU and and linear layer, having the latent feature dimension $16$. In all cases, we use $10$ Sinkhorn Operator iterations, with noise factor $1$ and a temperature of~$0.5$.  The output of the permutation network is a doubly stochastic matrix of dimension equal to the maximum node neighborhood size in the input graph. 

\subsection{Dataset details}
\label{app:datasets}

We use five datasets for evaluation:
\begin{enumerate}
\item \textbf{Twitter}~\cite{leskovec2012learning} is a snapshot of a part of Twitter's social network.%
\item \textbf{\google}~\cite{leskovec2010kronecker} is a snapshot of a part of Google-Plus social network.%
\item \textbf{Citeseer}~\cite{getoor2005link} is a snapshot of citation network.
\item \textbf{Cora}~\cite{getoor2005link} is a snapshot of citation network.
\item \textbf{PB}~\cite{ackland2005mapping} is a network of US political blogs.
    \end{enumerate}
Table~\ref{tab:datasets} shows some characteristics of the data sets we use.  They show a diversity of average degree, diameter, and number of node features. 

\begin{table}[hb]
\centering 
\maxsizebox{.9\hsize}{!}{
\begin{tabular}{|l||c|c|c|c|c|c|} \hline
Dataset &$|\Vcal|$&$|\Ecal|$& $d_{avg}$ & Diameter & \text{dim}$(\fb_{\bullet})$ & $|\Qcal|$  \\ \hline \hline
Twitter   & 193  & 7790  & 79.73 & 4  & 193  & 190 \\ \hline
\google   & 769  & 22515  & 57.56  & 7  & 769  & 718 \\ \hline
Citeseer   & 3312  & 7848  & 3.74  & 28  & 3703  & 1010 \\ \hline
Cora   & 2708  & 7986  & 4.90  & 19  & 1433  & 1470 \\ \hline
PB   & 1222  & 17936  & 28.36  & 8  & 1222  & 999 \\ \hline
\end{tabular} }
\caption{Dataset statistics.}
\label{tab:datasets}
\end{table}

\subsection{Discussion of evaluation protocols and metrics}
\label{sec:EvalMetrics}

As discussed in the main paper, we partition edges and non-edges into training, validation and test folds as follows.  Each query in the query node set $Q$ is required to be part of at least one triangle. For each $q \in Q$, in the original graph, we partition its neighbors $\nbr(q)$ and non-neighbors $\nnbr(q)$ into training, validation and test folds, where the corresponding node pairs are sampled uniformly at random.  In the main paper, these were in the ratio 54:6:40.  Here we also present results for the ratio  72:8;20.  We disclose the resulting sampled graph induced by the training and validation sets to the LP model.
After computing the scores for all potential edges, LP algorithms sort the potential edges in decreasing order of scores.
In this context, we note the following differences of our protocol from several prior works~\cite{ZhangC2018LinkPredGNN,hamilton2017inductive}.
\begin{enumerate}
    \item Some prior LP models, \eg, SEAL~\cite{ZhangC2018LinkPredGNN} remove a large fraction non-edges in the test set to ensure that the number of edges and non-edges in the test set is roughly equal. In contrast, we do not make
    any perturbation in the test set, which makes the evaluation more realistic as well as challenging. However,
    for completeness, we also present a comparative analysis of our method against the competitors by curating the test set to ensure that the number of edges and non-edges is roughly equal.
    \item Often in prior works~\cite{ZhangC2018LinkPredGNN,hamilton2017inductive}, the underlying LP algorithm sorts all potential edges $\ne$ by decreasing scores $\set{s(u,v)\,|\, (u,v)\in\ne}$ to output a single \emph{global} ranked list~$R$. However, in practical applications, no end-user (node) of the network observes the global ranking. Therefore, we assume that each node $q$ (regarded as a `query') is provided a \emph{local} ranking $R_q$ of recommended neighbors-to-be.  Recommending friends on Facebook, or movies on Netflix, are better served by this protocol.
\end{enumerate}
We measure the accuracy of an LP method in terms of Mean Average Precision (MAP) and Mean Reciprocal Rank (MRR), computed on the ranked lists of predicted neighbors across all the queries.  In particular, we compute: 
\begin{align}
\tMAP= \frac{1}{{|\Qcal|} } {\sum_{q\in \Qcal}} \text{AP}_q,\quad
\tMRR= \frac{1}{|\Qcal|} {\sum_{q\in \Qcal} }\frac{1}{r_q}, \label{eq:metrics}
\end{align}
where $\text{AP}_q$ is the average precision and $r_q$ is the rank of the topmost neighbor of the ranked list for the query node $q$.

\subsection{Hyperparameters and policy parameters}

For all training, we impose an early stopping criteria based on validation fold AUC and AP scores.  We remember the performance from the latest 100 epochs (the so-called `patience' parameter).  If the relative variation in AUC and AP fall before the fraction $10^{-4}$, we stop training and roll back to the best model in the patience window.

We train our LP model using the ranking loss defined in Eqn~\eqref{eq:HingeRankingLoss} with choices of optimizer, learning rate and margin as summarized in Table~\ref{tab:PermGnnHparams} for reproducibility.

The hashing network $C_\psi$ is trained according to the loss defined in Eqn.~\eqref{eq:HashOpt}, with the hyperparameters $\alpha$ and $\beta$ set to 0.01 for all datasets. 
In all cases, we train the network $C_\psi$ using SGD optimizer with learning rate of~0.05.

\begin{table}[ht]
\centering 
\maxsizebox{.8\hsize}{!}{
\begin{tabular}{|l||c|c|c|} \hline
Dataset &Learning Rate& Margin & Optimizer \\ \hline \hline
Twitter   & $5{\times}10^{-4}$  & 0.01 & Adam  \\ \hline
\google   & $5{\times}10^{-5}$  & 0.01 & SGD \\ \hline
Citeseer   & $5{\times}10^{-5}$ & 0.1 & SGD \\ \hline
Cora   & $5{\times}10^{-5}$ & 0.1 & SGD \\ \hline
PB   & $5{\times}10^{-6}$ & 0.01 & SGD \\ \hline
\end{tabular} }
\caption{Dataset specific hyperparameters of \our.}
\label{tab:PermGnnHparams}
\end{table}


\end{document}